# Comprehensive Analysis and Experimental Design of High-Gain DC-DC Boost Converter Topologies

**Webster Adepoju[1,]* Mary Sanyaolu[2,]***

[1] Department of Electrical and Computer Engineering, Tennessee Technological University, Cookeville, United States; webster4live@gmail.com

[2] Department of Earth Sciences, Tennessee Technological University, Cookeville, United States; omadepoju22@gmail.com

*Correspondence webster4live@gmail.com

**Abstract**

Global demand for clean and eco-friendly energy sources has inspired decades of far-reaching research in power generation from renewable energy sources. Solar cells, wind, and tidal sources are limited in output power generation compared to the fast-rising power requirements of most industrial applications. Besides, the efficiency of conventional DC-DC boost converters is significantly low due to the presence of parasitic elements culminating in switching losses. This study presents three high-gain boost converter topologies for optimizing the limited voltage generation by solar Photovoltaic cells. The three converters are realized based on a modification of a classical Cuk converter. Simply put, the first proposed converter is realized by the inclusion of one capacitor and one inductor to the classical topology. Similarly, the addition of 3 capacitors, 2 diodes, and one inductor leads to the practical realization of the second proposed topology. Similarly, the third proposed topology consists of additional 3 diodes and 3 capacitors. Based on this estimation, the first, second, and third proposed high gain modified Cuk converter topologies generate output voltages 10 times, 20 times, and 29 times the input voltage respectively when the switching device is gated at 90% duty ratio. Theoretical/mathematical analysis validates the Lt-spice numerical simulation results of all the proposed converters. Furthermore, experimental prototype results were compared with Lt-spice estimation to determine the accuracy of the converters.

## 1. Introduction

The quest for a healthy, safe, and eco-friendly environment has been a major driver of cutting-edge research in efficient energy generation. The most common means of power generation are nuclear plants, fossil fuels, hydroelectric, coal, etc. However, these conventional energy sources generate by-products that pose unprecedented levels of health hazards to the environment. This is evident in climate change and global warming, conditions responsible for the continuous depletion of the ozone layer and exposure of the ecosystem to vast amounts of Ultraviolet (UV) radiation.

Conversely, renewable energy sources, including solar, wind, tidal, etc. are cheaper and inexhaustible compared to conventional energy sources [1]. However, efficient power harvesting from renewable energy sources is often hampered by myriads of factors such as fluctuating irradiation, unpredictable weather conditions, and partial shadow. System reliability and efficient power harvesting are some of the issues that need to be overcome to fully maximize the utilization of renewable energy sources, especially solar systems [1]-[3]. The above drawbacks can be solved by feeding the generated voltage from the solar panel into a high-gain boost converter [3-6]. The voltage conversion ratio, M(k) of a conventional boost converter is comparatively small and the output voltage is equally insufficient to meet the rising voltage demands of high-power applications. High efficiency and steep voltage gain are requirements for the utilization of renewable energy. Over the years, different voltage boost DC-DC converters ranging from high power to low power (HVDC transmission) have been decimated in literature. Among these are: multilevel and multistage converters [7-11]; interleaved designs [11-13], switched capacitors /inductors type [14], transformer-less topologies [14, 15], Cuk converter [16, 17], Single Ended Primary Inductor Converter (SEPIC) [18, 19]; Luo converters [20, 21] to mention a few. It is worth noting that each of these topologies has its unique advantages and drawbacks. Extreme duty ratio often results in increased losses and reduced efficiency. This paper presents three modified Cuk converter topologies having a steep voltage turn ratio. Several years of research in power electronics have culminated in the development of numerous voltage-boosting techniques for DC-DC converters. These include Switched capacitors [23, 24], Magnetic coupling [26], Voltage Multiplier network [25, 26], and Voltage lift -Switched Inductor (VL-SL) [27].

The boosting technique of the three proposed modified Cuk circuits presented in this study is based on the Voltage Lift Technique (VLT). Various voltage lift methods have been reviewed and compared in the literature [20, 21]. The Cuk circuit is an inverting buck-boost converter and it has the flexibility of generating output voltage lower or higher than the input depending on the duty ratio, k. Minimal ripples and increased efficiency make the Cuk converter preferable to other conventional topologies. Low ripples are attributable to the presence of an output low pass filter as depicted in Fig. 1. The converter's output capacitor is usually a large electrolytic capacitor having low Equivalent Series Resistance (ESR).

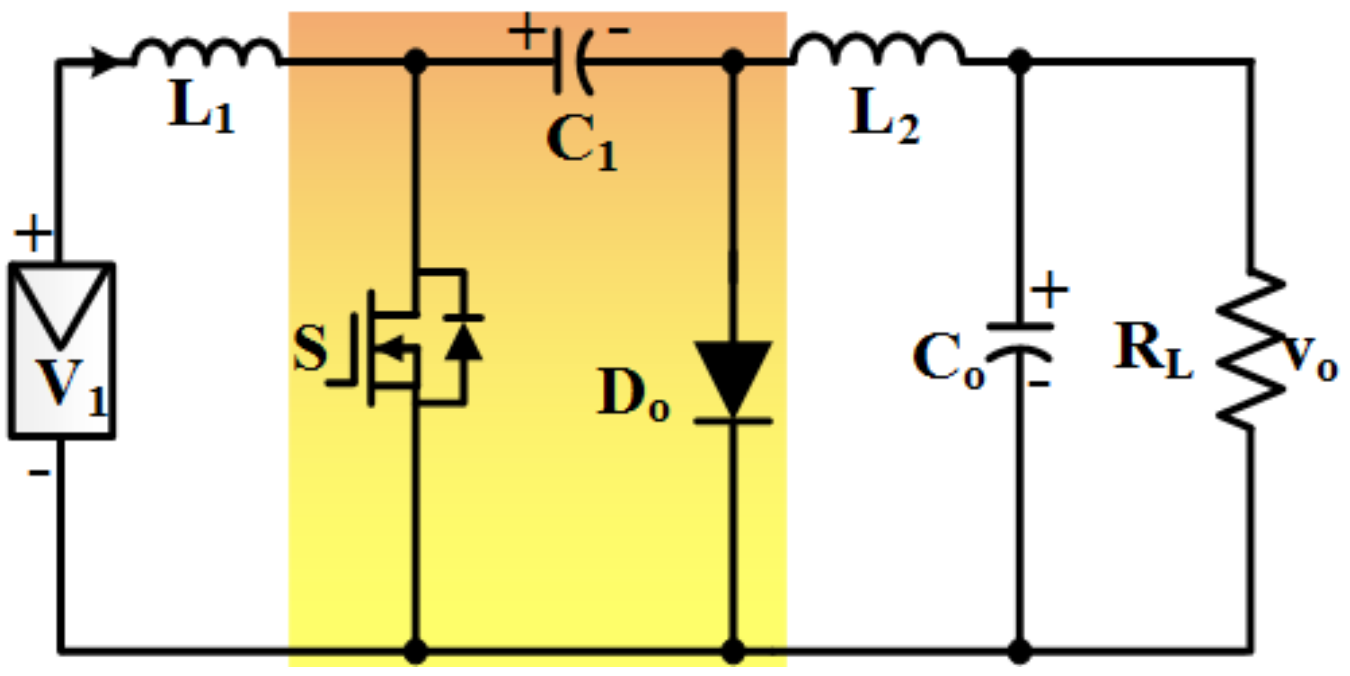

**Figure. 1:** Classical Cuk DC-DC Converter.

## 2. High gain modified Cuk Converter – first proposed circuit

Fig. 1 and Fig. 2 depict the circuit diagram of the classical Cuk converter and the proposed high gain modified (boost mode) Cuk converter respectively. The proposed circuit comprises a MOSFET switch S, and inductive components $L_1$, $L_2$. $L_3$, capacitive components $C_1$, $C_2$, $C_3$, resistive load $R_L$ and an ideal diode $D_o$. The input capacitor $C_1$ was chosen to be a large electrolytic capacitor that has a low ripple current rating and a low Equivalent Series Resistance (ESR) to ensure the converter operates with maximum efficiency at a steady state.

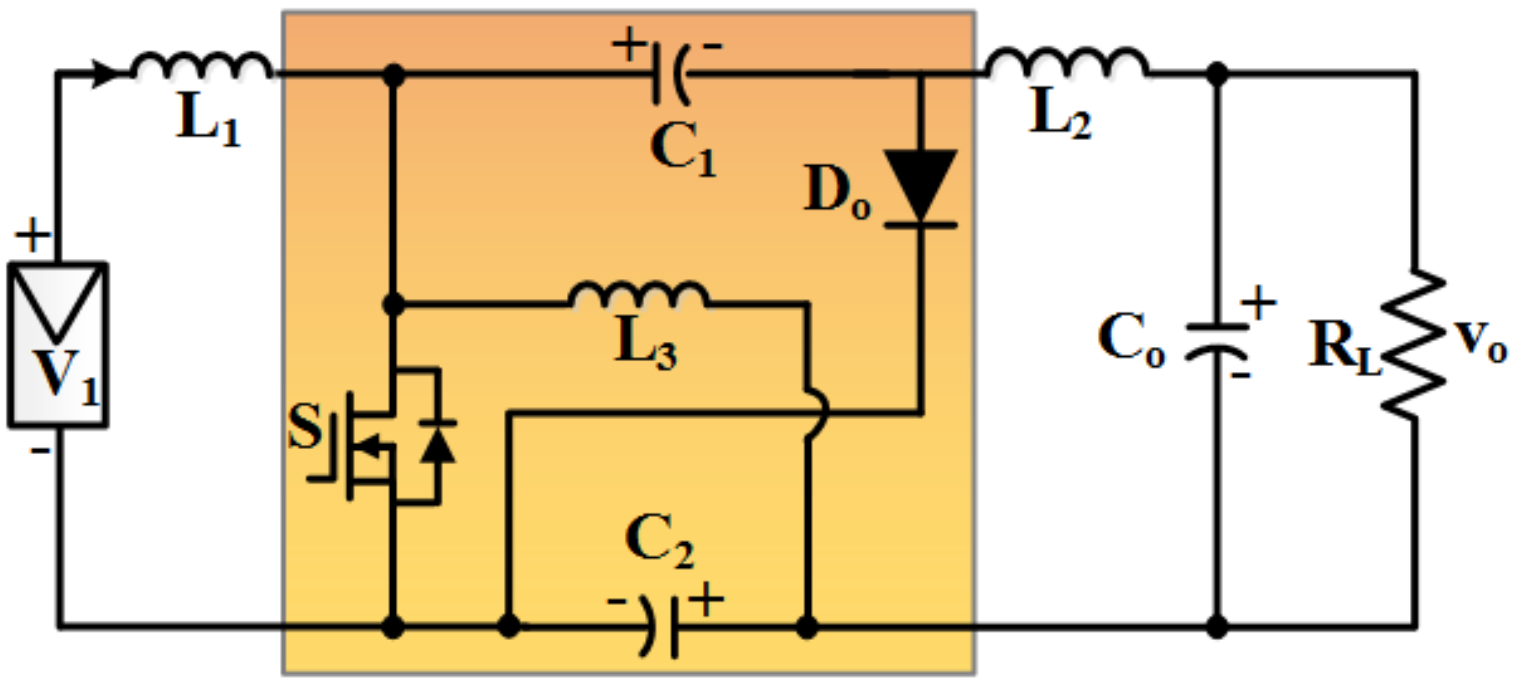

**Figure. 2:** Circuit Configuration of the first proposed Converter.

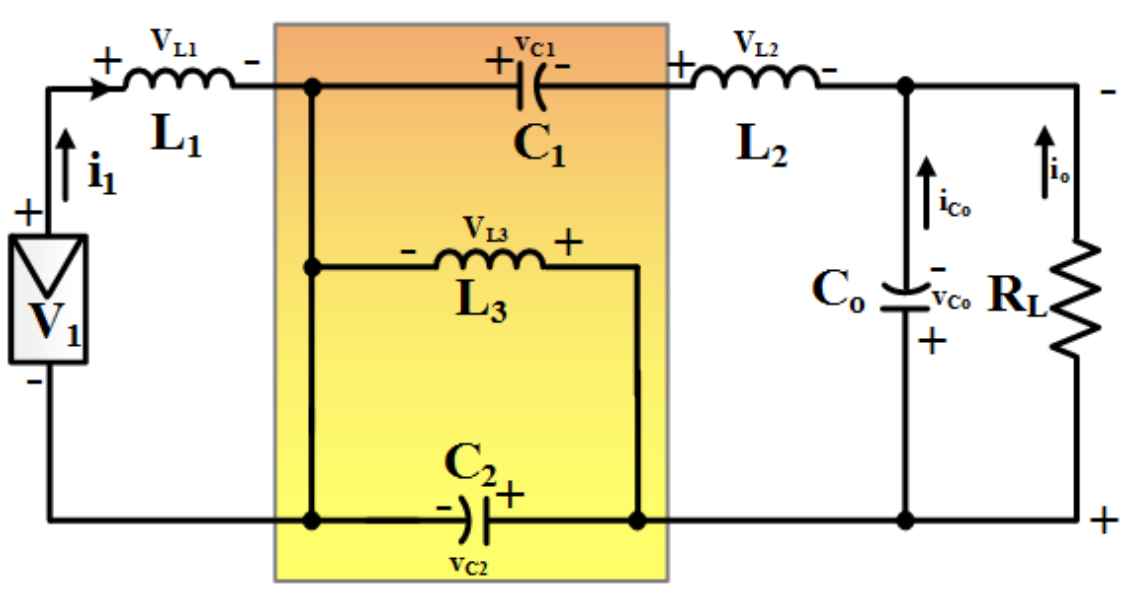

**Figure.3:** ON state of the proposed circuit

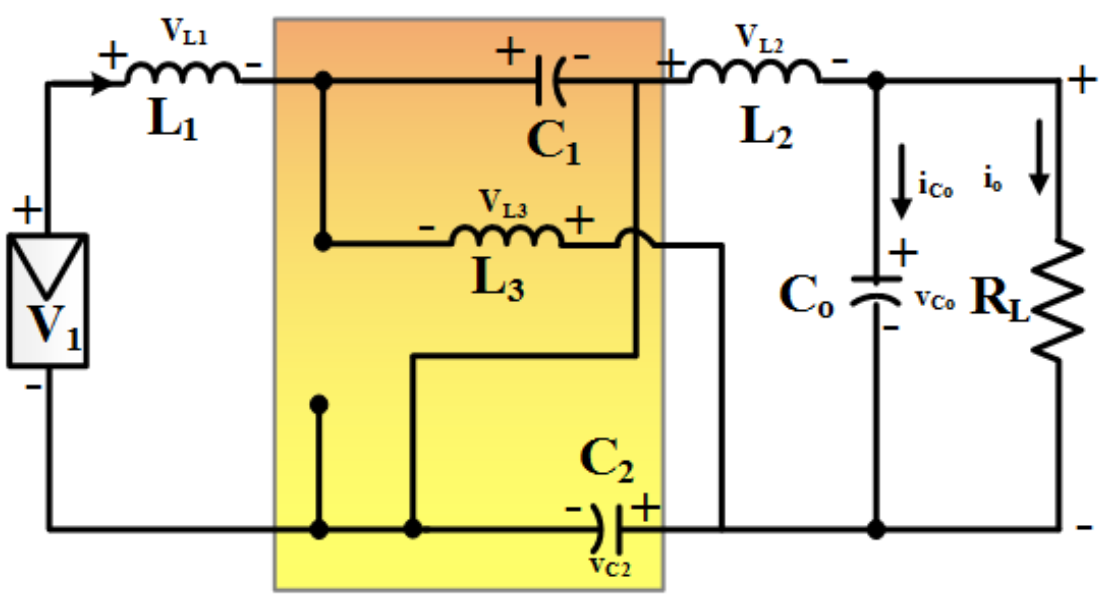

**Figure. 4:** OFF state of the proposed circuit

At steady state condition and assuming the switch S is in the ON state, then equations (1)–(3) are valid for the circuit in Fig. 3 based on Kirchhoff's Voltage Law (KVL),

Assuming the switch S is OFF, then the following expressions hold true for inductors $L_1$, $L_2$ and $L_3$;

$$V_{L1} = V_1 \tag{1}$$

$$V_{L2} = V_O - V_{C2} - V_{C1} \tag{2}$$

$$V_{L3} = - V_{C3} \tag{3}$$

When the static switch is OFF, equations (4)-(6) describe the circuit according to KVL.

$$V_{L1} = V_1 - V_{C1} \tag{4}$$

$$V_{L2} = -V_O - V_{C2} \tag{5}$$

$$V_{L3} = V_{C1} + V_{C2} \tag{6}$$

Using the inductor Volts second balance principle and assuming steady state operation, the mean voltage flowing through an inductor is zero. Hence, equation (7) depicts the steady state behavior of the proposed modified Cuk circuit

$$kV_{LON} = (1-k)V_{LOFF} \tag{7}$$

Using the expression in (7), the steady equations for $L_1$, $L_2$ and $L_3$ can be written thus;

For $L_1$;

$$kV_1 = (1-k)(V_1 - V_{C1}) \tag{8}$$

$$(2k-1) = (k - 1)V_{C1} \tag{9}$$

For $L_2$;

$$k(Vo-V_{C2}-V_{C1}) = (1-k)(-Vo + V_{C2}) \quad (10)$$

$$Vo(2k-1) = V_{C2} + kV_{C1} \quad (11)$$

For $L_3$;

$$-kV_{C3} = (1-k)\,(V_{C1} + V_{C2}) \quad (12)$$

$$V_{C2} = (1-k)\,V_{C1} \quad (13)$$

Combining equation (9), (11) and (13) gives the duty ratio of the converter. Similarly, when both (11) and (13) are combined, we have the expression:

$$V_o = (2k-1)V_{C1} \quad (14)$$

If equation (14) is inserted into equation (9), then the expression for the voltage gain of the converter can be deduced as;

$$\frac{Vo}{V_I} = -\frac{1}{1-k} \quad (15)$$

The experimental prototype and Lt-spice simulation parameter of the proposed circuit are presented in Table 1. In addition, the prototype design specification is depicted in Table 2. Similarly, the experimental prototype design of the proposed circuit showing the signal generator, digital oscilloscope, and power supply is shown in Figure 5.

**Table 1:** Part Number for Experimental Design.

| COMPONENTS | PART NUMBER AND VALUE |
| --- | --- |
| Switch S | IRF630(MOSFET) |
| Inductors $L_1$; $L_2$ ;$L_3$ | High current radial inductors |
| Capacitor $C_0$; $C_1$; $C_2$; $C_3$;$C_4$ | Cap electrolytic 10µF 250V ST |
| Diode Do; D1, D2, D3. | MBR1060 schottky diode |
| Load $R_1$ | 1000Ω /25W resistor |
| Digital oscilloscope | TBS1000 |
| Signal generator | Wavetek 3001 |
| DC power supply | YH-305D |

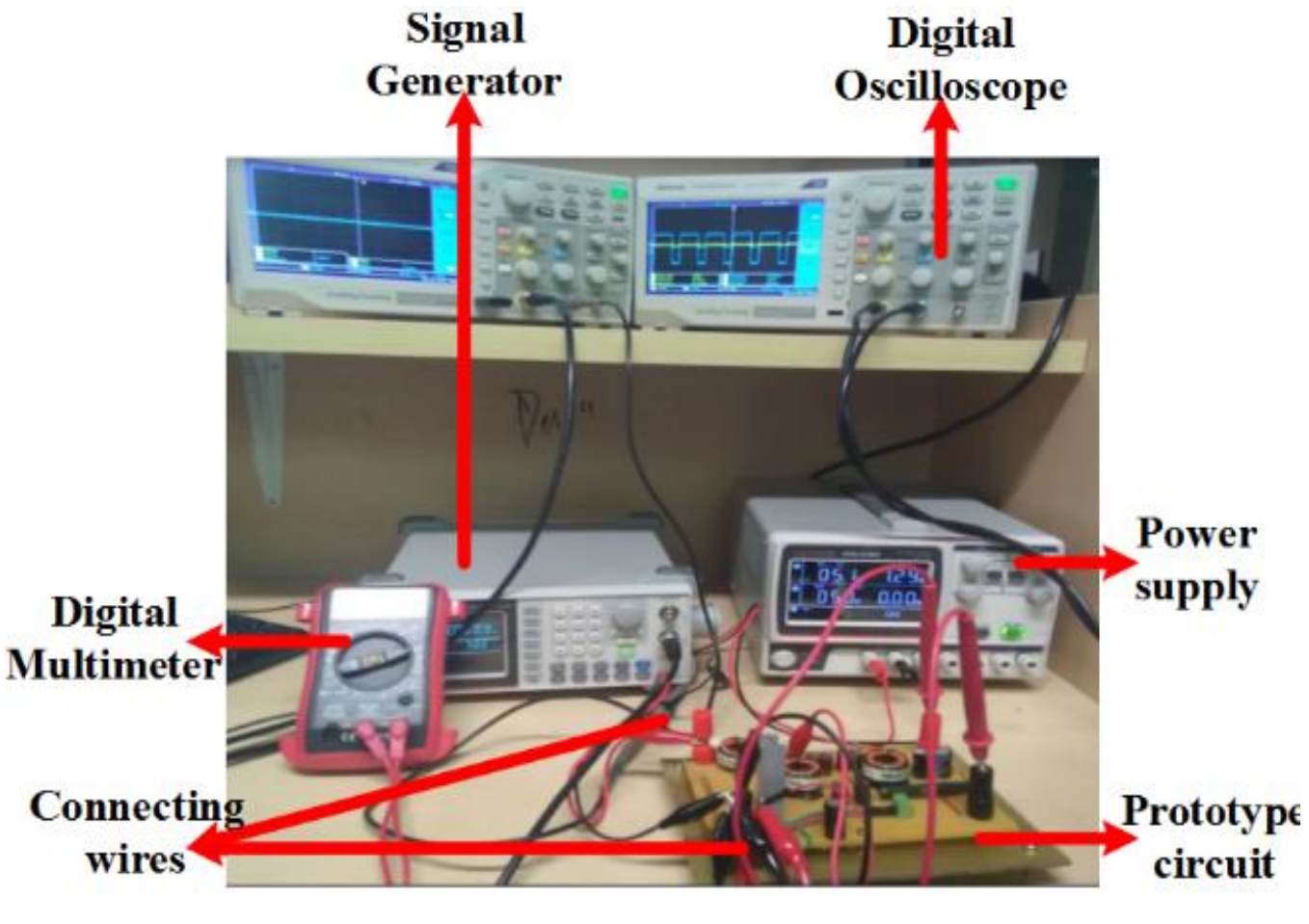

**Figure 5:** Experimental Set-up of the first proposed circuit.

**Table 2:** Prototype/Lt-spice Design Specifications

| PARAMETER | VALUE |
|---|---|
| Input voltage | 5V |
| Output power | 1.25Watt |
| Switching frequency | 120kHz |
| Output voltage | 25V |
| Resistor | 100Ω |

## 3. Experimental Prototype and Lt-Spice Simulation Results.

5V DC voltage is fed from the power supply and a square wave (Pulse Width Modulation, PWM) from the signal generator is supplied to the gate of the MOSFET. The signal generator is set to a switching frequency of 120 kHz. The duty ratio is varied step-wisely from 10%-90% and afterward, the corresponding value of the output voltage is measured for each variation in the duty ratios.

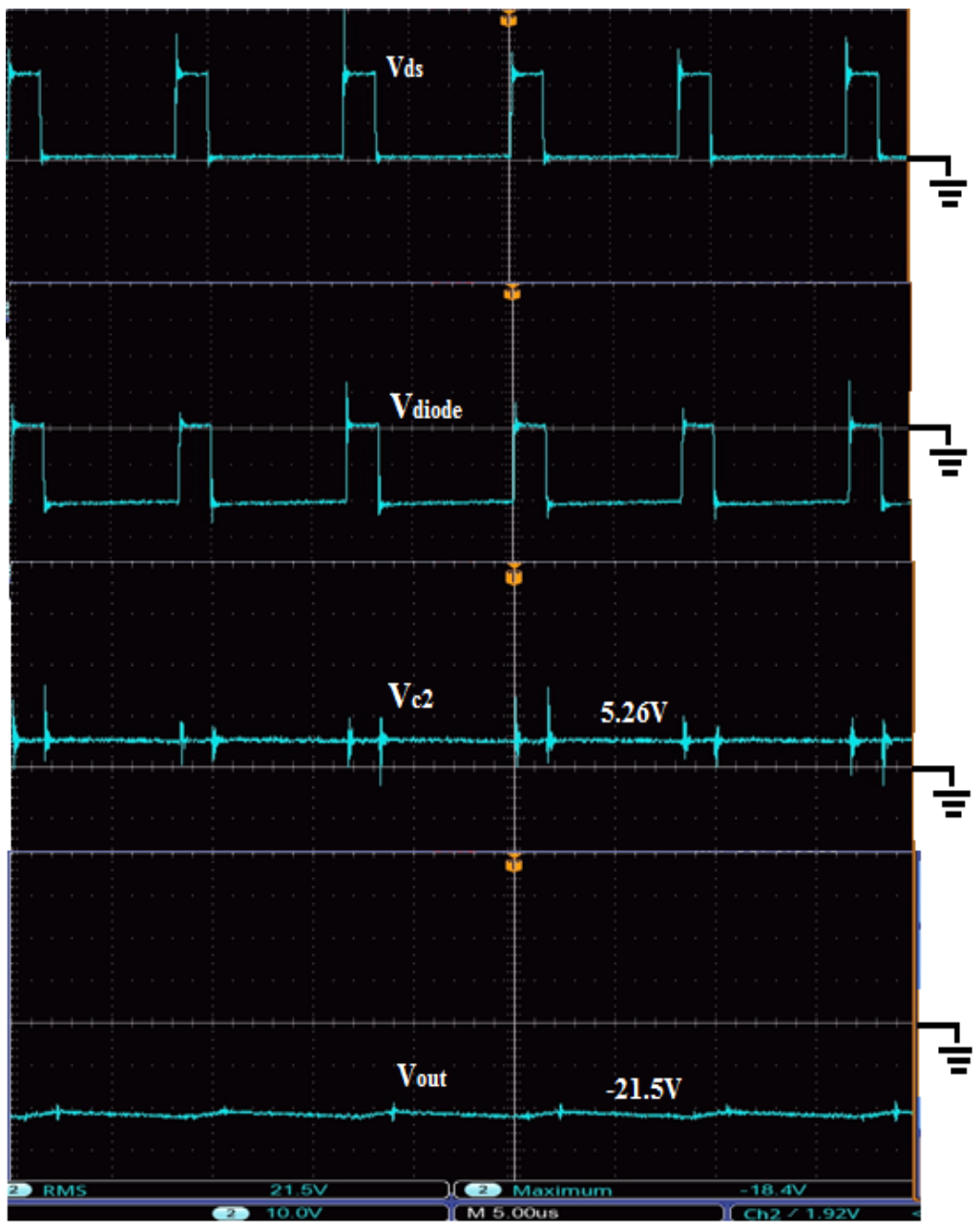

**Figure 6:** Experimental prototype waveforms

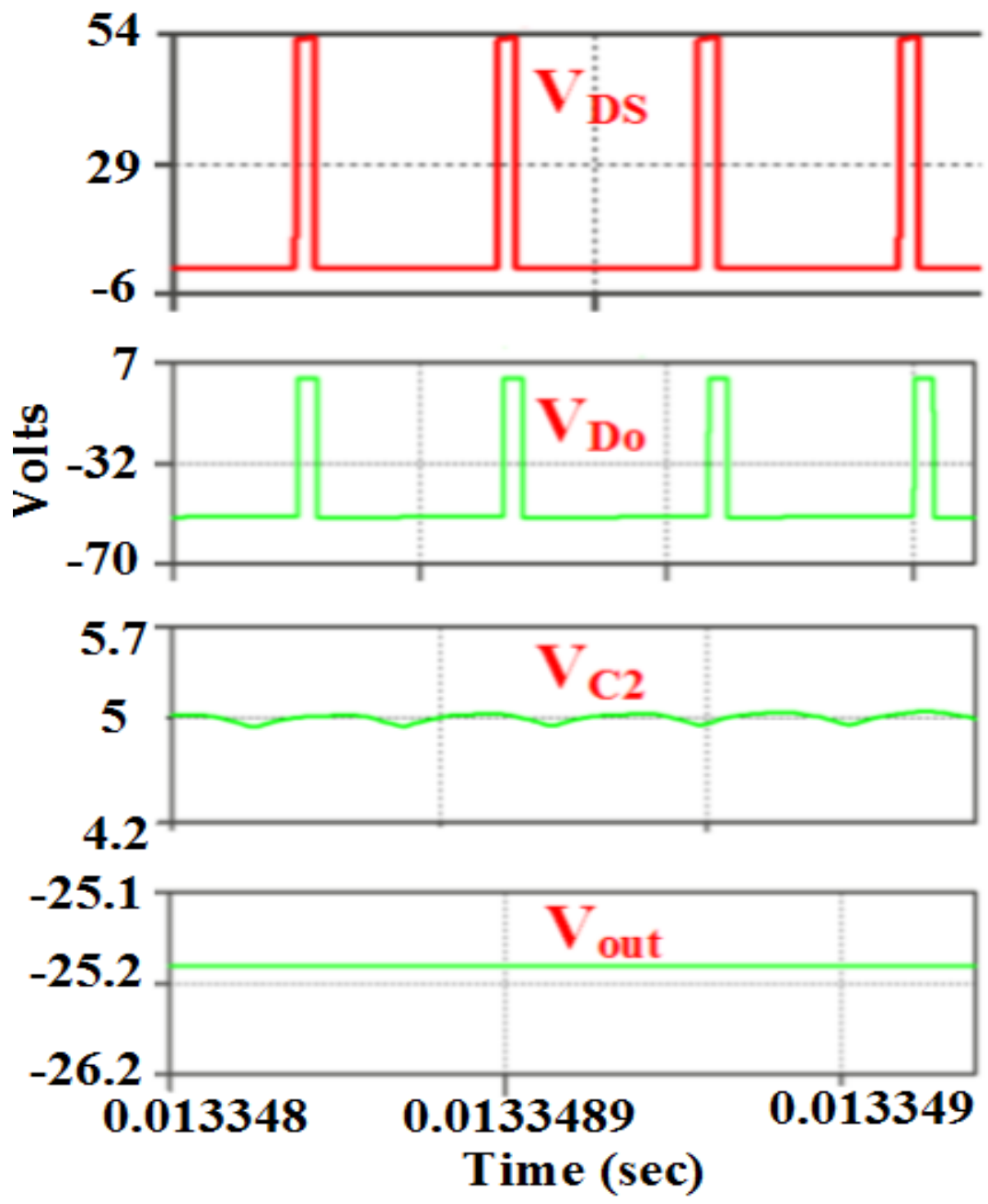

**Figure 7:** Lt-spice simulation waveforms

Figure 6 and Figure 7 above respectively illustrate the TBS1000 experimental prototype waveforms and Lt-spice numerical simulation behavior of the proposed circuits given a 5V input voltage and 80% duty ratio. From Figure 7 & Figure 8, when switch S is on, $V_{GS}$ is high and $V_{DS}$ is low, the diode is off and capacitor $C_1$ is discharged by inductor $L_2$ and similarly, capacitor $C_2$ is charged by inductor $L_3$. However, the diode conducts the currents on inductors $L_1$ and $L_2$ when the switch is off. Inductor $L_3$ and capacitor $C_1$ are charged by the current from inductor $L_1$. The voltage spike on the switch is caused by the high number of reactive components, leading to an increase in the current stress on the switching device. With an input voltage of 5V and 80% duty ratio, the converter generates an output voltage of 21.5V, equivalent to a deviation of 3.5V from the theoretical estimation.

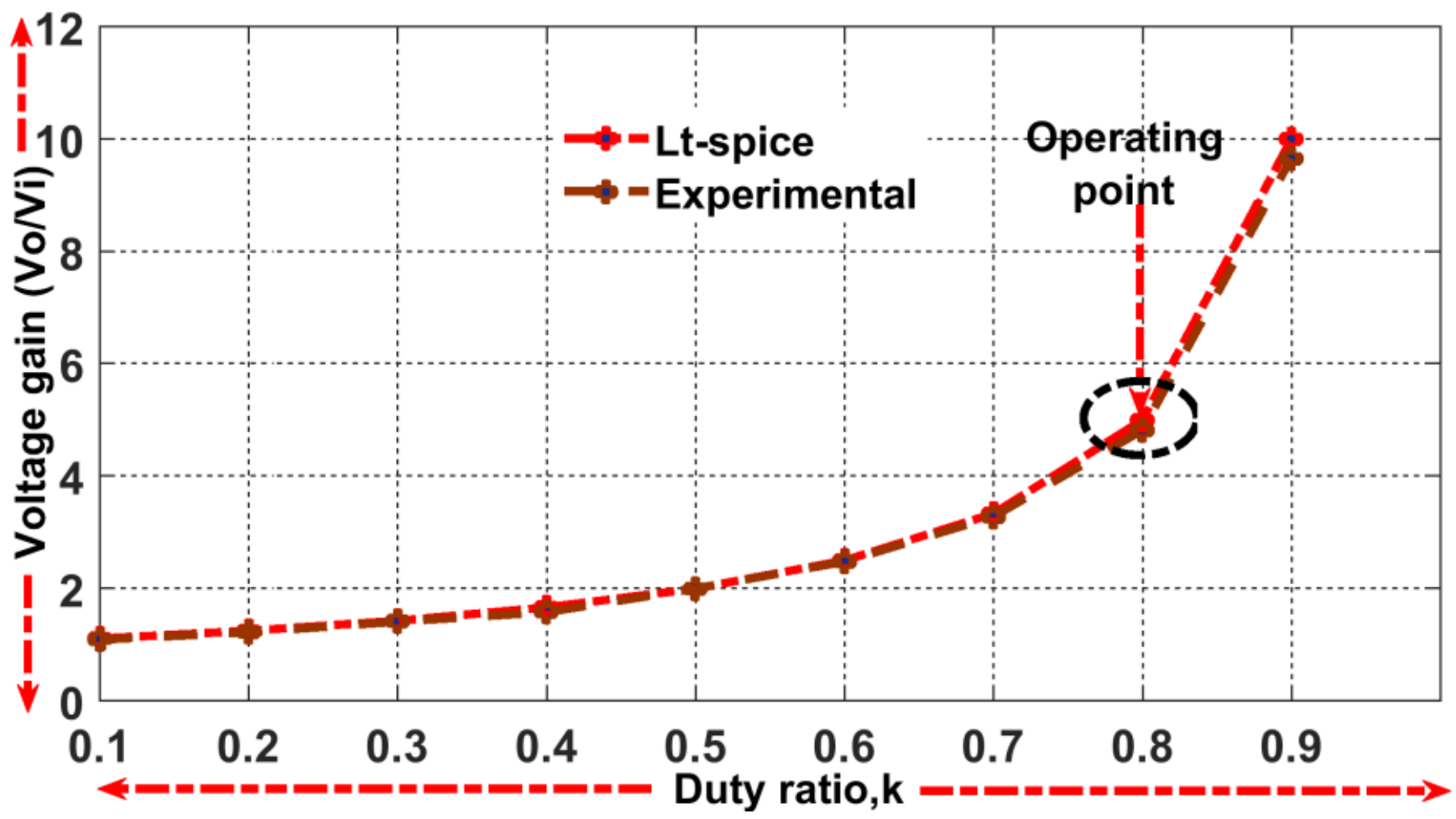

**Figure 8**: Lt-spice and experimental voltage gain of the proposed circuit.

Furthermore, Fig. 8 above illustrates the Lt-spice numerical simulation and experimental prototype voltage gain plot. The curve clearly shows that the prototype waveforms closely match the simulation results, further validating the accuracy of the theoretical analysis. The switching components in the prototype waveform experience high voltage spikes due to parasitic inductance generated by the wiring in the circuitry. This overshoot often causes the switching devices to malfunction if their magnitudes are excessive. Spikes are common in power converters and can be reduced by incorporating a snubber circuit.

## 4. High Gain Cuk Converter - Second Proposed Topology.

The second proposed circuit is an extension of the first proposed modified Cuk circuit, the difference being the inclusion of a voltage multiplier cell. Figure 9 below illustrates the pictorial diagram of the proposed circuit. In terms of components count, the converter comprises 3 diodes, 5 capacitors, and 3 inductors. The proposed converter consists of two sections denoted by I and II. Section I is a switched capacitor network while section II is a voltage boost multiplier cell. Each of the segments constitutes a voltage gain of -1/1-k, where k is the duty ratio of the gate drive. The ON state and OFF state equivalent circuits of the proposed converter are depicted in Figure 10 and Figure 11 respectively.

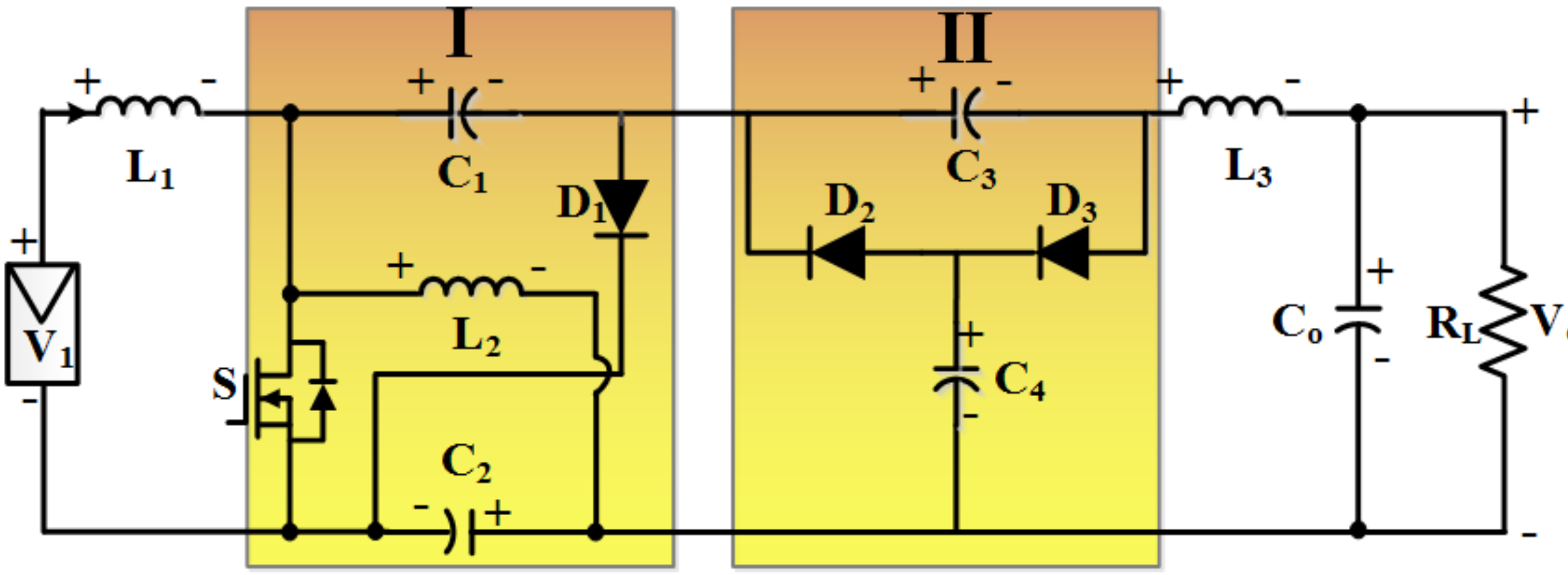

**Figure 9:** Circuit configuration of second proposed high gain modified Cuk converter.

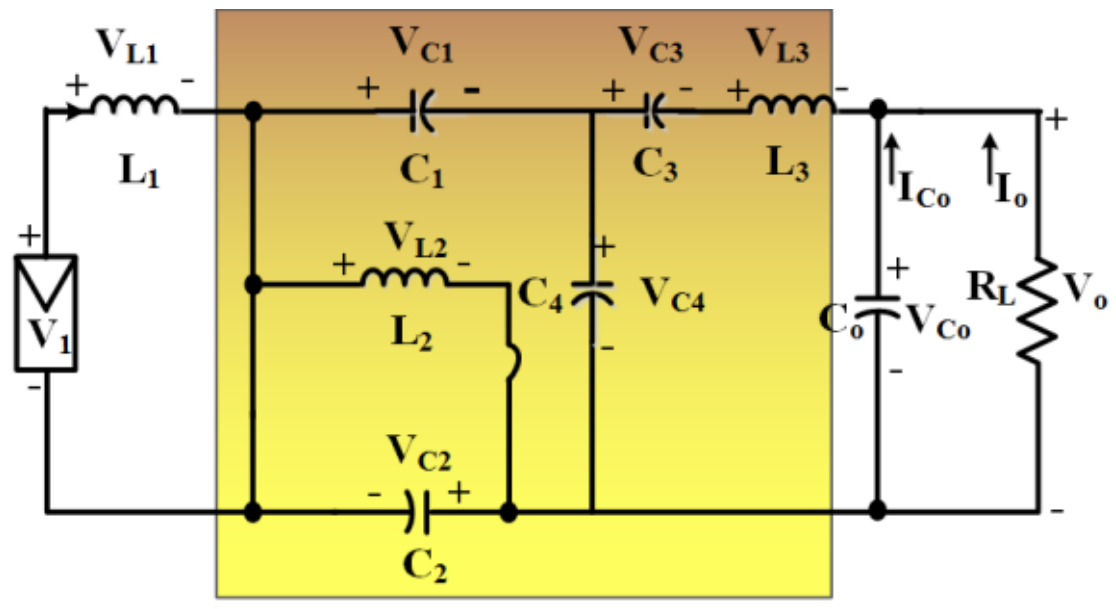

**Figure 10:** ON state equivalent circuit.

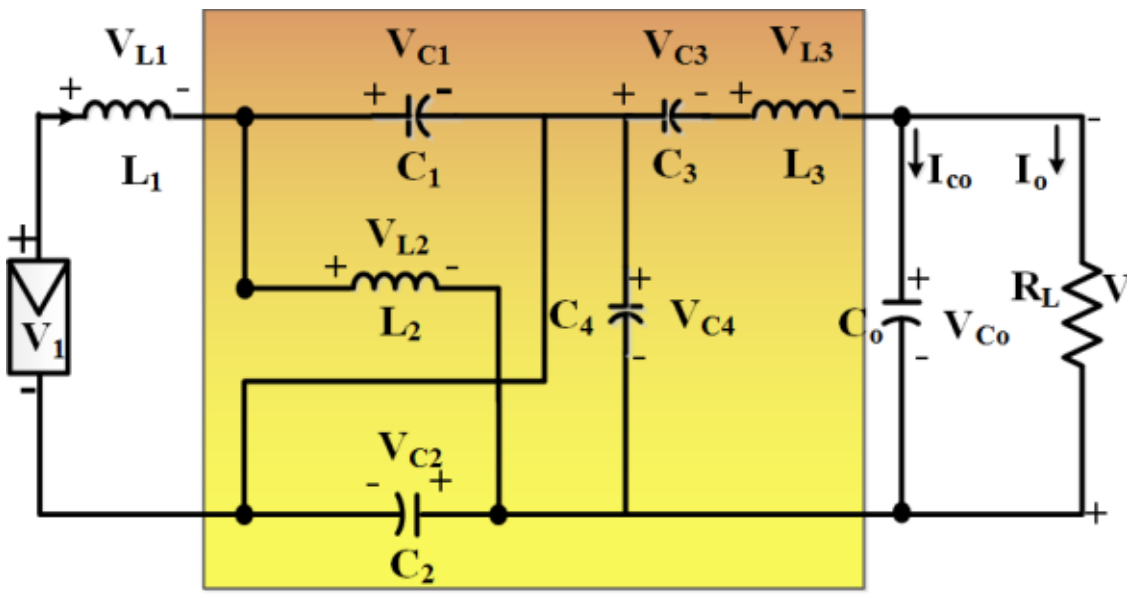

**Figure 11**: OFF state Equivalent Circuit.

Theoretical investigation shows that the overall voltage transformation ratio of the proposed circuit is 20V/V, corresponding to a 90% operating duty cycle. Theoretical analysis is done using Faraday inductor volts sec balance equation shown in (1). The switch ON and OFF equations for inductors $L_1$, $L_2$, and $L_3$ are obtained thus:

Considering Inductor $L_1$;

When the switch is ON;

$$V_{L1} = V_1, \tag{16}$$

When the switch is OFF;

$$V_{L1} = V_1 - V_{C1}, \tag{17}$$

Using inductor volt second balance principle in (1) yields;

$$kV_1 = (1-k)(V_1 - V_{C1}) \tag{18}$$

When (18) is simplified, it gives;

$$V_{C1} = \frac{(2k-1)V_1}{(k-1)} \tag{19}$$

For Inductor $L_2$;

When the switch is ON;

$$V_{L2} = V_{C2} = V_{C1} + V_{C4} \tag{20}$$

When the Switch is OFF, equations (21)-(22) hold true;

$$V_{L2} = V_{C2} - V_{C1} \tag{21}$$

$$V_{C2} = V_{C3} + V_{C4} \tag{22}$$

Using Faraday Law in (1) and inserting (21) and (22) yields equation (23)-(25);

$$kV_{C2} = (1-k)(V_{C2} - V_{C1}) \tag{23}$$

$$kV_{C2} = V_{C2} - V_{C1} - kV_{C2} + kV_{C1} \tag{24}$$

$$V_{C2}(2k-1) = V_{C1}(k-1) \tag{25}$$

Inserting (19) into (25) gives (26)-(27);

$$V_{C2} = \frac{k-1}{2k-1}\frac{(2k-1)V_1}{(k-1)} \tag{26}$$

$$V_{C2} = V_1 \tag{27}$$

Inserting (22) into (21), and using Faraday Law in (1) yields (28)-(30);

$$k(V_{C1} + V_{C4}) = (1-k)(V_{C3} + V_{C4} - V_{C1}) \tag{28}$$

$$kV_{C1} + kV_{C4} = V_{C3} + V_{C4} - V_{C1} - kV_{C3} - kV_{C4} + kV_{C1} \tag{29}$$

$$V_{C4}(2k-1) = V_{C3}(1-k) - V_{C1} \tag{30}$$

Different notation of Faraday Law in (1) yields the following equation.

$$k(V_{C1} + V_{C4}) = (1-k)(V_{C2} - V_{C1}) \tag{31}$$

$$V_{C4} = \frac{V_{C2}(1-k) - V_{C1}}{k} \tag{32}$$

If (27) is inserted into (32);

$$V_{C4} = \frac{V_1(1-k) - \frac{(2k-1)}{(k-1)}V_1}{k} \tag{33}$$

This simplifies to;

$$V_{C4} = \frac{-k}{(k-1)}V_1 \tag{34}$$

If (34) is inserted into (30);

$$\frac{-k}{(k-1)}V_1\ (2k\text{-}1)\ )+\frac{(2k-1)V_1}{(k-1)}= -V_{C3}\ (1\text{-}k) \tag{36}$$

$$\frac{(2k-1)V_1(1-k)}{(k-1)} = -V_{C3}\ (1\text{-}k) \tag{37}$$

$$\frac{-k}{(k-1)}V_1\ (2k\text{-}1)= -V_{C3}\ (1\text{-}k)\text{-}\ \frac{(2k-1)V_1}{(k-1)} \tag{35}$$

This simplifies to;

$$V_{C3} =\frac{(2k-1)V_1}{(1-k)} \tag{38}$$

This simplifies to;

$$V_{C3} =\frac{(2k-1)V_1}{(1-k)} \tag{39}$$

For Inductor $L_3$;

When the switch is ON;

$$V_{L3}=V_{C3}+V_{C4}+V_{Co} \tag{40}$$

When the switch is OFF;

$$V_{L3}=V_{C4}+V_{Co} \tag{41}$$

If Faraday Law in (1) is used for $L_3$;

$$k(V_{C3}+V_{C4}+V_{Co})=(1\text{-}k)(V_{C4}+V_{Co}) \tag{42}$$

This simplifies to;

$$kV_{C3}+kV_{C4}+ kV_{Co}=V_{C4}+V_{Co}\text{-}\ kV_{C4}\text{-}kV_{Co} \tag{43}$$

$$kV_{C3}+V_{C4}(2k\text{-}1)+V_{Co}(2k\text{-}1)=0 \tag{44}$$

If (16) is inserted into (29);

$$kV_{C3}+ V_{C3}\ (1\text{-}k)\text{-}V_{C1} +V_{Co}(2k\text{-}1)=0 \tag{45}$$

This simplifies to;

$$V_{C3}+ V_{Co}\ (2k\text{-}1)=V_{C1} \tag{46}$$

Inserting (5) and (24) into (31) yields;

$$\frac{(2k-1)V_1}{(1-k)}+ V_{Co}\ (2k\text{-}1)=\frac{(2k-1)V_1}{(k-1)} \tag{47}$$

Finally, turn ratio of the proposed circuit can be found;

$$\frac{V_{Co}}{V_1} = -\frac{2}{(1-k)} \quad (48)$$

Lt-spice numerical simulation and hardware prototype design of the proposed circuit are examined to validate the correctness of the theoretical equations.

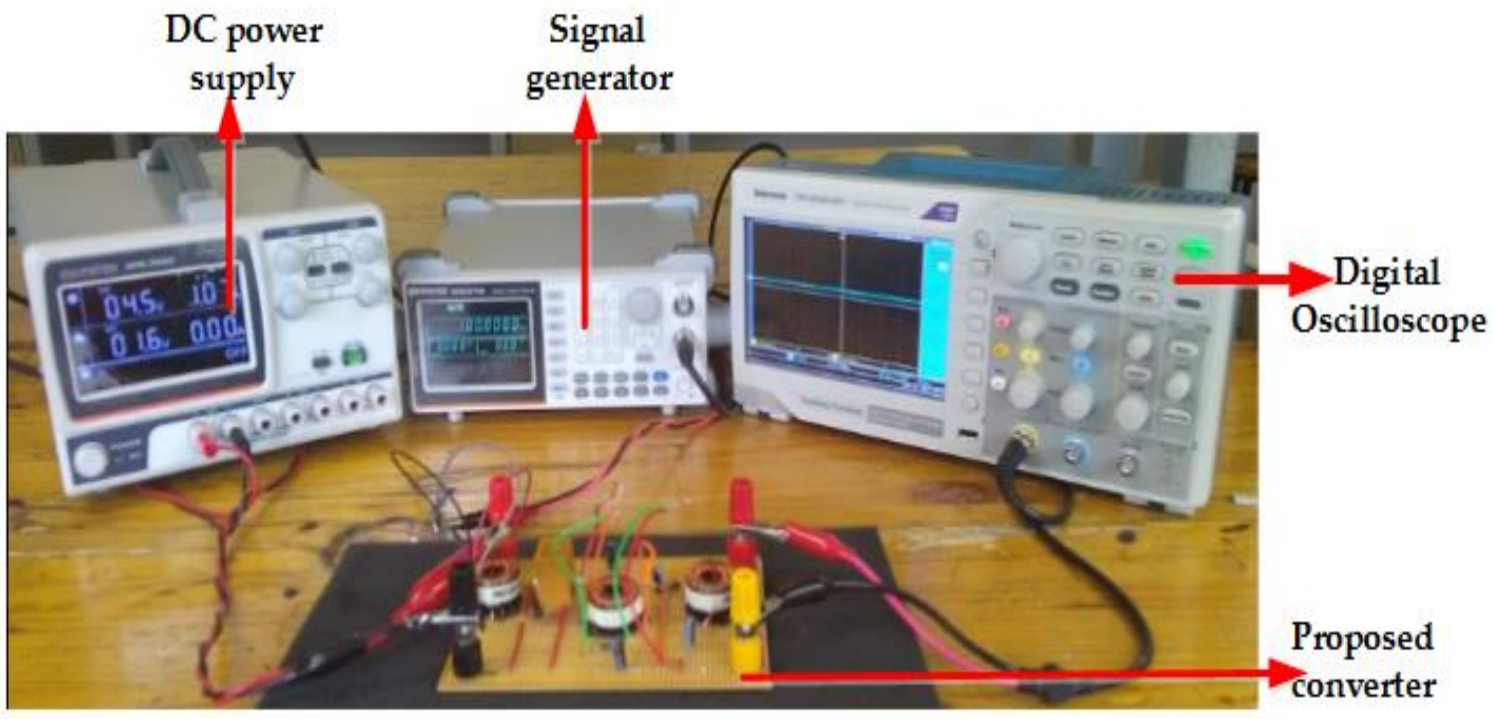

**Figure 12**: Experimental prototype set-up of the proposed circuit.

**Table 3**: Prototype/Lt-spice Design Specifications of the Proposed Modified Circuit

| PARAMETER | VALUE |
|---|---|
| input voltage | 5V |
| Output power | 2.5Watt |
| Switching frequency | 120kHz |
| Output voltage | 50V |
| Resistor | 100Ω |

The experimental set-up of the proposed circuit is demonstrated in Figure 12 above. Moreover, the component parameter values for both Lt-spice simulation and prototype design is the same as in Table 1 above. In addition, Table 3 above depicts the parameter design specification for the modified circuit. Going forward, a 5V DC voltage is fed into the converter from the power supply as shown in Figure 12. PWM is supplied to the MOSFET gate at a switching frequency of 120 kHz while slowly varying the duty cycle from 10% to 90%. Corresponding waveforms of the capacitors and diodes were obtained from TBS1000 digital oscilloscope. Prototype/simulation waveforms of the capacitors and diode components of the proposed converter are depicted in shown in Figures (13-21) below.

## 5. Experimental/prototype and Lt-spice simulation results of second proposed converter.

On the Lt-spice simulation results, the green color represent the capacitor and diode waveforms while red color denotes the MOSFET gate signal, $V_{Gs}$. In the same way, the yellow lines on the prototype waveforms represent the PWM signal while the blue waveforms illustrate the diode and capacitor voltages. As shown in Figure 13(a) & (b), when the MOSFET gate signal, VGs is ON and VDs is OFF. Similarly, when VGs is OFF, VDs is ON, thus proving that the MOSFET is switched on. The spikes associated with the VDs prototype waveform is due to switching losses in the MOSFET.

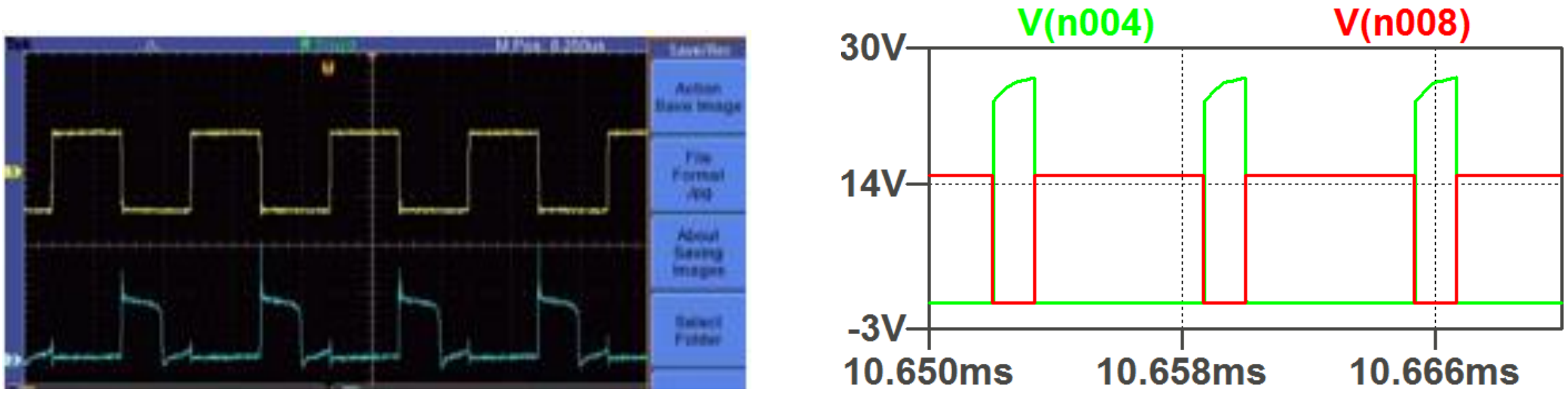

**(a)** **(b)**

**Figure 13:** $V_{GS}$ and $V_{DS}$ (a) prototype waveform (b) Lt-spice waveform.

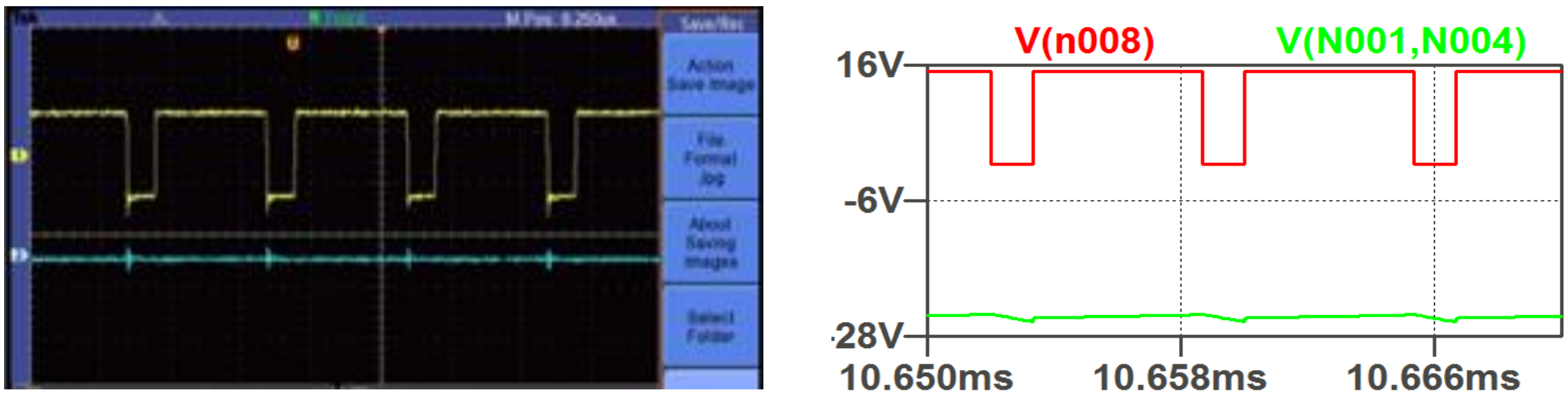

**(a)** **(b)**

**Figure 14**: Capacitor voltage, $V_{C1}$ (a) prototype waveform (b) Lt-spice waveform

.

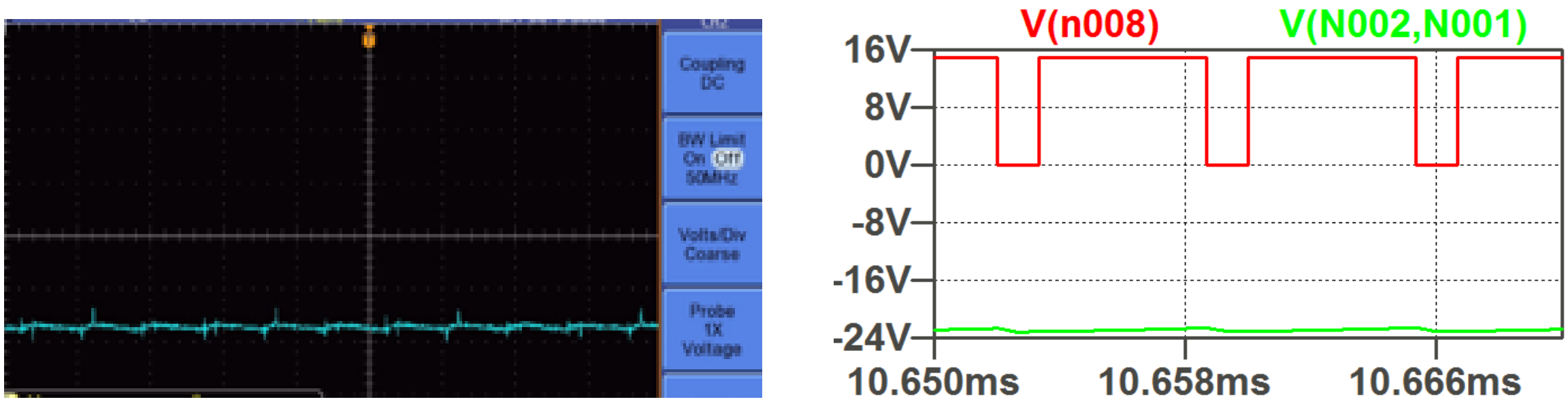

**(a)** **(b)**

**Figure 15**: Capacitor voltage, $V_{Co}$ (a) experimental result (b) Lt-spice simulation.

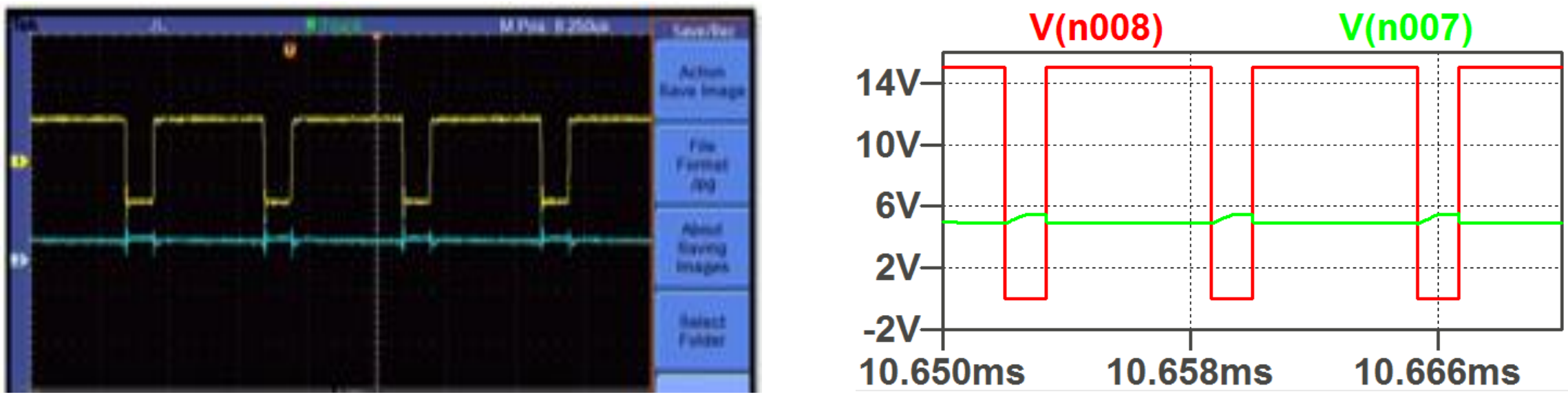

**(a)** **(b)**

**Figure 16:** Capacitor voltage, $V_{c2}$ (a) experimental result (b) Lt-spice simulation.

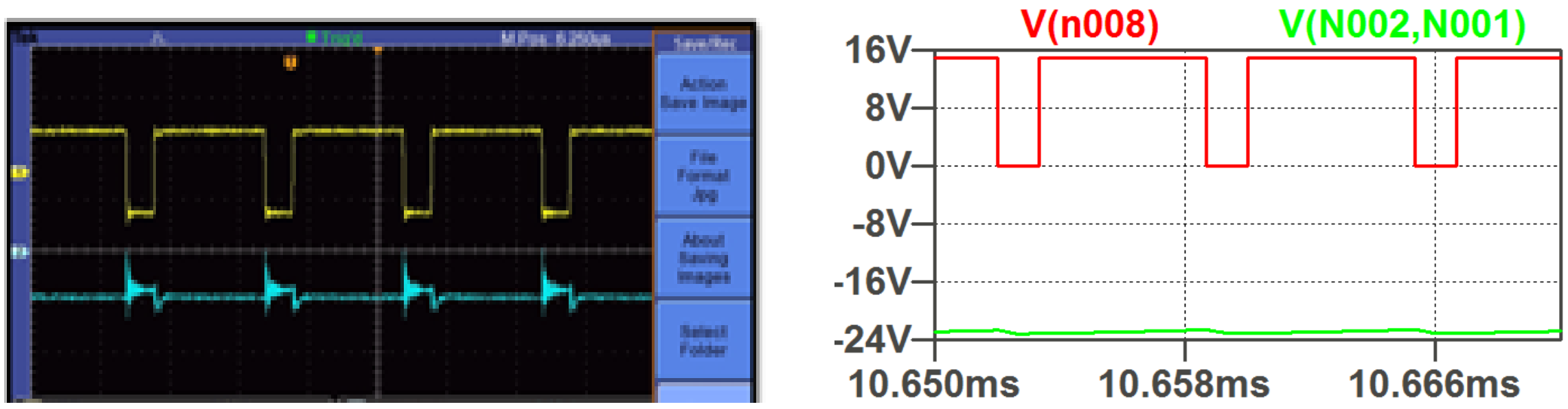

(a) (b)

**Figure 17:** Capacitor Voltage, $V_{c3}$ e (a) prototype waveform (b) Lt-spice waveform

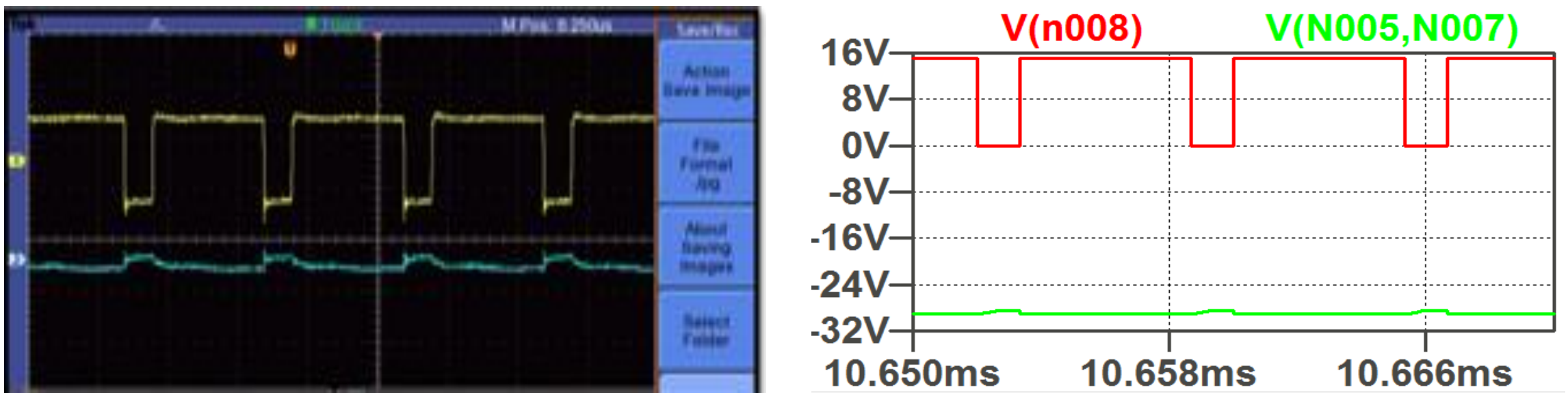

(a) (b)

**Figure 18:** Capacitor voltage, $V_{C4}$ (a) prototype waveform (b) Lt-spice waveform.

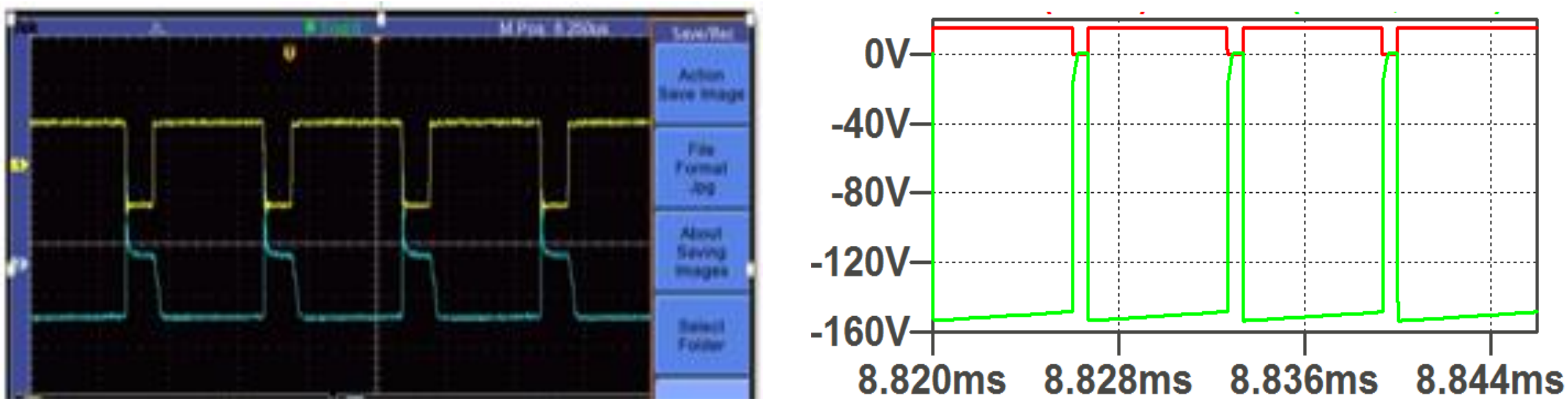

(a) (b)

**Figure 19:** Diode Voltage, $V_{D1}$ (a) prototype waveform (b) Lt-spice waveform

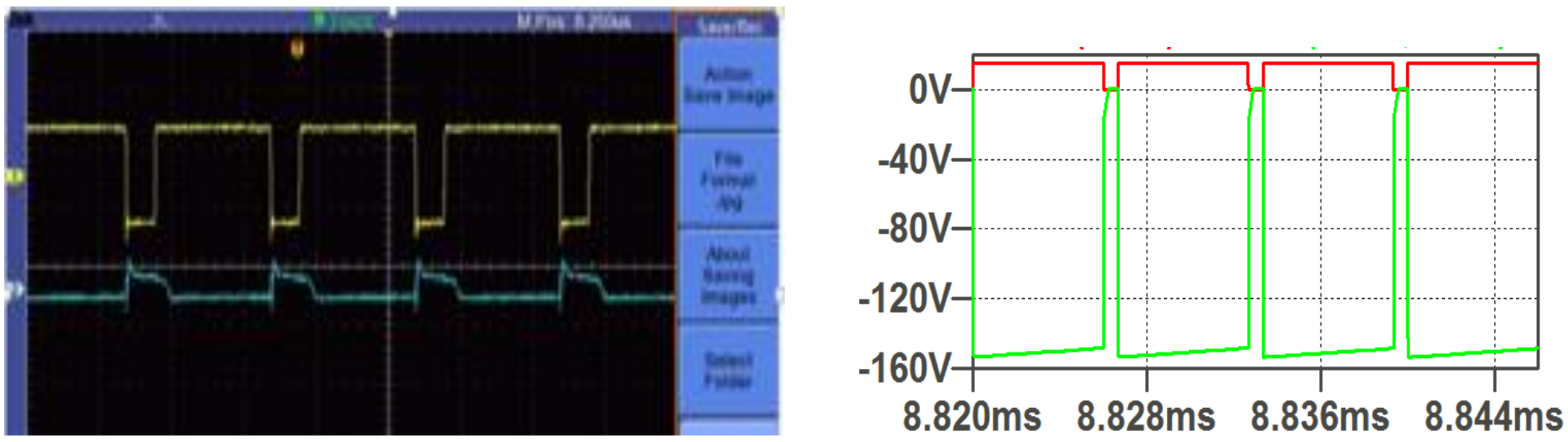

(a) (b)

**Figure 20:** Diode Voltage, $V_{Do}$ (a) prototype waveform (b) Lt-spice waveform.

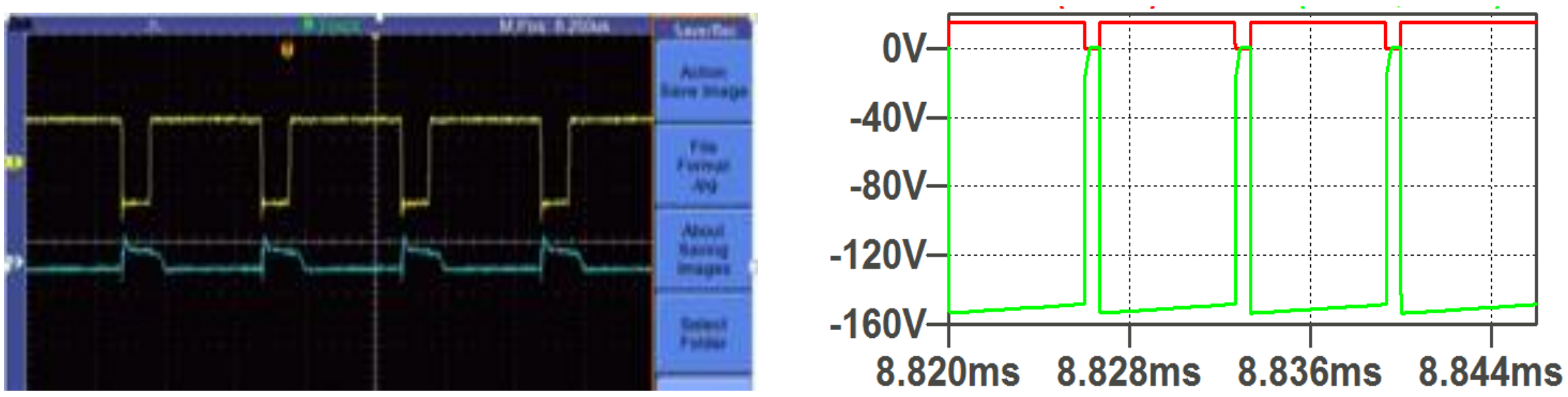

**(a)** **(b)**

**Figure 21:** Diode D2 Voltage (a) prototype waveform (b) Lt-spice waveform

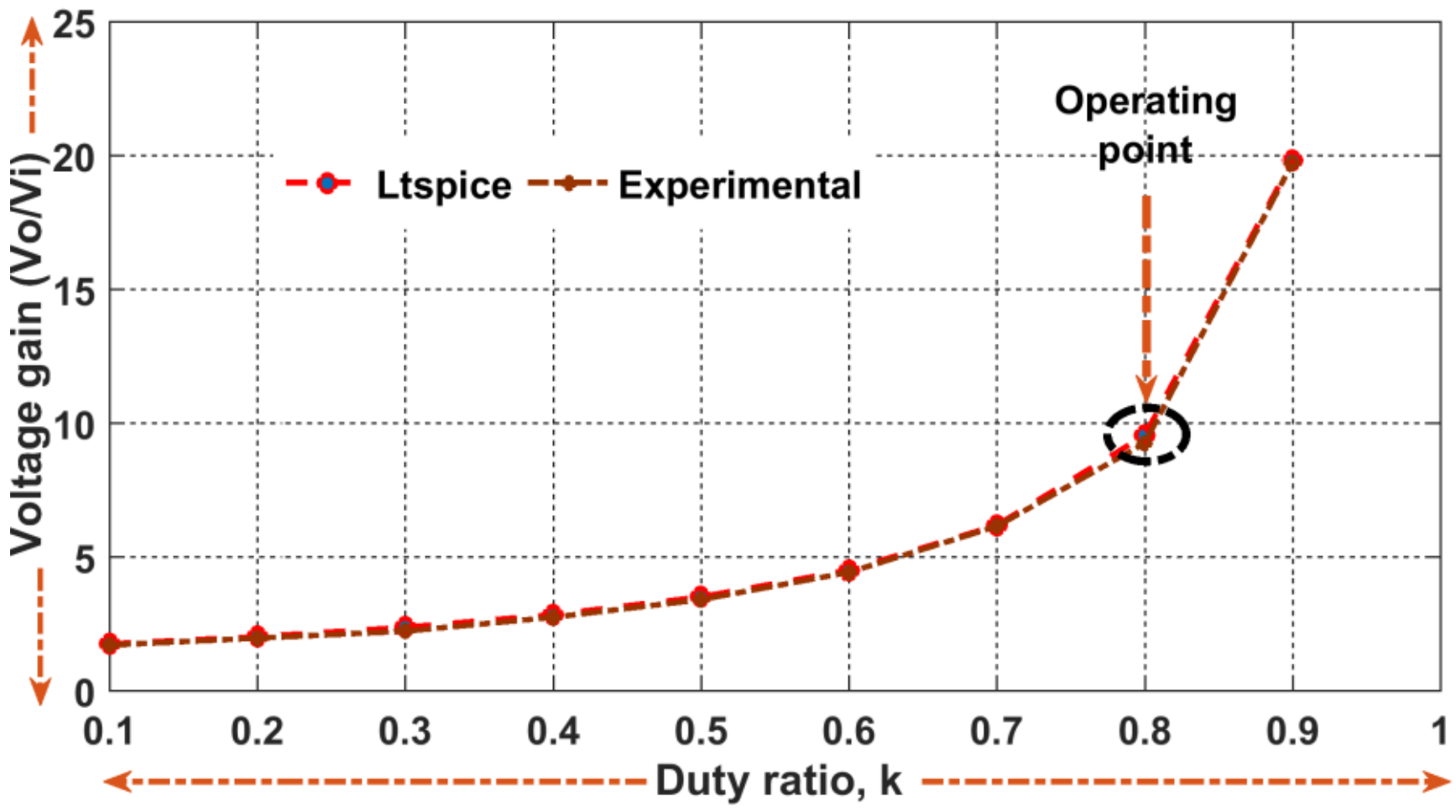

**Figure 22:** Prototype /Lt-spice voltage gain comparison plot of the proposed circuit

The experimental and Lt-spice voltage gain comparison plot of the proposed converter is depicted in Figure 22 above. From the curve, it is obvious that both hardware results and Lt-spice investigation give acceptable results. In addition, the obtained performance indices closely match the theoretical analysis thus confirming the accuracy of the converter. However, the practical efficiency of the converter dropped appreciably at a very high duty ratio as a result of a large voltage drop. High duty cycle degrades the switch performance and also increases the switching losses.

## 6. High Output Gain Modified Ćuk Converter - Third Proposed Topology

The third proposed circuit is illustrated in Fig. 23 below. The circuit is subdivided into two segments symbolized by **I** and **II**. A total of 2 inductors, 5 capacitors, and 4 diodes constitute the component count of the high-gain modified Cuk converter.

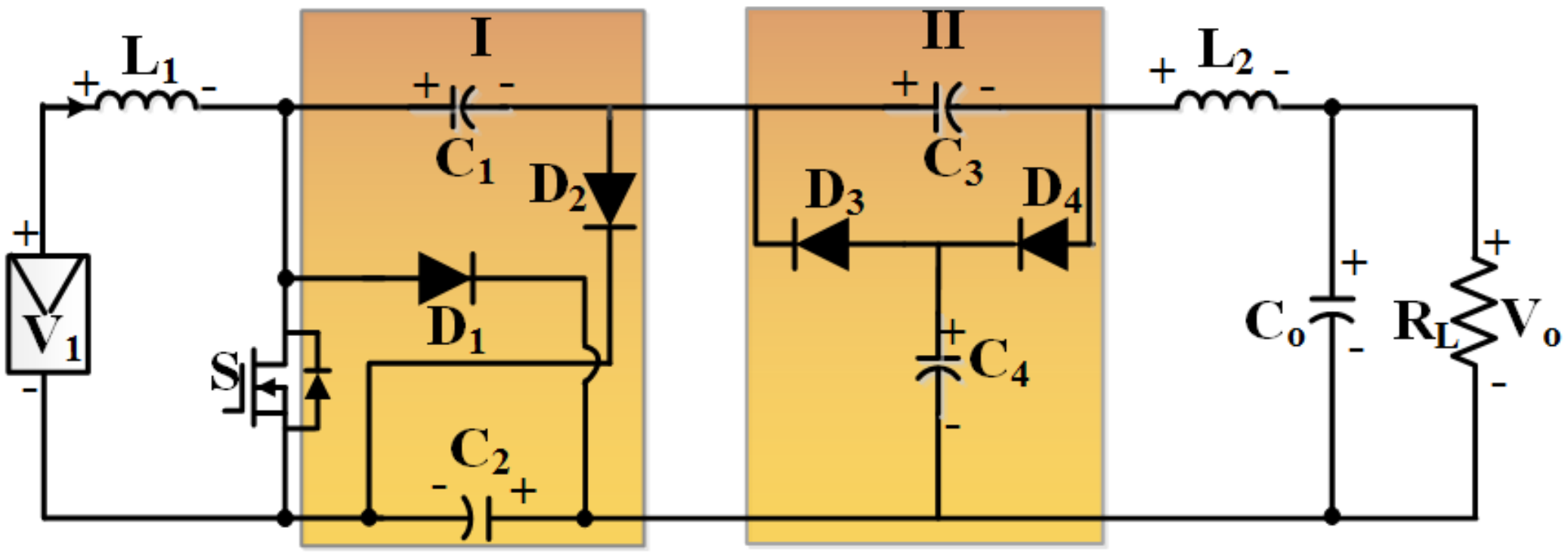

**Figure 23.** Topology of modified Ćuk DC-DC converter

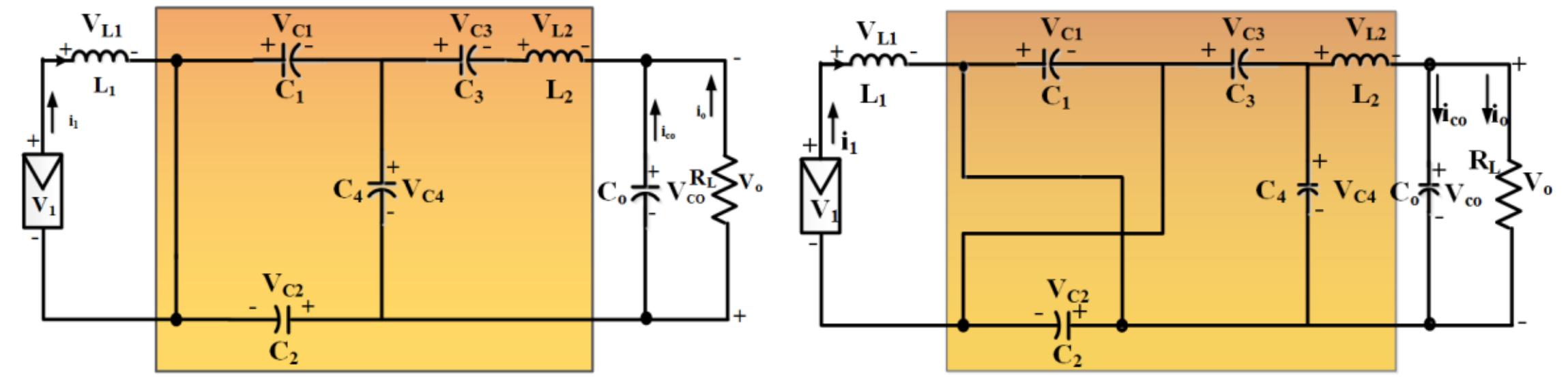

(a) (b

**Figure 24:** Proposed Circuit (a) ON state Equivalent Circuit (b) OFF state Equivalent Circuit

The circuit achieves an overall theoretical voltage fain of (2+k)/ (1-k) at peak duty ratio. This implies the converter's output voltage is 29 times greater than the input voltage if the duty ratio (k) is 0.9. The voltage turns ratio of segment I is (1+k)/ (1-k) [26], while segment II accomplishes 1/ (1-k). Thus, the overall voltage gain of the proposed modified Cuk circuit is obtained by a simple algebraic summation of the voltage turns ratio of segment I and segment II. The ON state and OFF state configurations of the modified circuit is demonstrated in Fig. 24 (a) & (b) respectively. Theoretical investigation of the converter's voltage turns ratio can be deduced by using the inductor volts sec balance principle depicted in (49).

$$kV_{L(ON)} = (1 - k)V_{L(OFF)} \tag{49}$$

When the switch S is turned ON and OFF, the following equations can be written;

$L_1$ over one period for steady state conditions.

When S is ON;

$$V_{L1} = V_1, \tag{50}$$

$$V_{L1} = V_{C1} - V_1, \text{ If S is OFF} \tag{51}$$

$$kV_1 = (1 - k)(V_{C1} - V_1) \tag{52}$$

The expression in (4) given in can be simplified as;

$$V_{C1} = \frac{(2k-1)V_1}{(1-k)} \tag{53}$$

Similarly, the following expressions holds true for inductor $L_2$ in steady state operation;
When S is ON;

$$V_{L2} = V_{C3} + V_{C4} + V_{Co} \tag{54}$$

$$V_{C4} = V_{C1} + V_{C2} \tag{55}$$

When S is OFF;

$$V_{L2} = -V_{C4} + V_{Co} \tag{56}$$

$$V_{C1} = V_{C2} \tag{57}$$

$$V_{C4} = -(V_{C3} + V_{C2}) \tag{58}$$

Faraday Law given in (49) can be applied to different combinations of above equations. First, If (53) and (56) is inserted into (49), following equation can be obtained.

$$kV_{C3} + kV_{C4} + kV_{Co} = (1-k)(-V_{C4} + V_{Co}) \tag{59}$$

$$kV_{C3} + V_{Co}(2k-1) = -V_{C4} \tag{60}$$

Second, (7) is inserted into (6), and (10) is inserted into (8), Faraday law can be written as follows.

$$k(V_{C3} + V_{C1} + V_{C2} + V_{Co}) = (1-k)(V_{C3} + V_{C2} + V_{Co}) \tag{61}$$

(61) can be simplified as follows;

$$(2k-1)(V_{C3} + V_{C2} + V_{C0}) + kV_{C1} = 0 \tag{62}$$

Similarly, following equation can be written by using (6), (8) and (10)

$$k(V_{C3} + V_{C4} + V_{Co}) = (1-k)(V_{C3} + V_{C2} + V_{Co}) \tag{63}$$

(63) can be simplified as follows.

$$(2k-1)(V_{C3} + V_{C0}) + kV_{C4} = V_{C2}(1-k) \tag{64}$$

Inserting (62) into (64) simplifies a valuable equation for the analytical analysis of the circuit, and one can verify that the following equation holds for steady state.

$$V_{C4} = V_{C1} + V_{C2} \tag{65}$$

Third, $V_{C1} = V_{C2}$ equation can be proved by writing the following equation

$$k(V_{C3} + V_{C4} + V_{Co}) = (1-k)(V_{C1} + V_{C3} + V_{Co}) \tag{66}$$

Similarly,

$$k(V_{C3} + V_{C4} + V_{Co}) = (1-k)(V_{C2} + V_{C3} + V_{Co}) \tag{67}$$

Thus, it can be stated from (66) and (67) that $V_{C1} = V_{C2}$ holds at steady state. The following equation can be written by using (55), (56) and (63).

$$k(V_{C3} + 2V_{C1} + V_{Co}) = (1-k)(V_{C1} + V_{C3} + V_{Co}) \tag{68}$$

(68) can be simplified as follows;

$$(2k-1)(2V_{C1} + V_{Co}) + kV_{C3} = 0 \tag{69}$$

If (62) and (69) are written in the same equations, it can be concluded from polynomial equations properties that $V_{C1} = V_{C2} = V_{C3}$. Finally, (60) can be simplified as follows.

$$kV_{C1} + V_{Co}(2k-1) = -2V_{C1} \tag{70}$$

$$V_{Co}(2k-1) = -V_{C1}(2+k) \tag{71}$$

Inserting (53) into (68) yields;

$$V_{co}(2k-1) = -\frac{(2k-1)(2+k)}{(1-k)} V_1 \tag{24}$$

The desired equation could be obtained.

$$\frac{V_{co}}{V_1} = -\frac{2+k}{1-k} \tag{25}$$

## 7. Experimental Prototype/ Lt-spice Simulation Results and Discussion.

First, the circuit is implemented on a Lt-spice simulation environment to validate the correctness of the theoretical equations. Furthermore, the experimental prototype design of the converter is constructed and tested with a 5V DC supply voltage as shown in Fig.25. Gate signal is supplied to the MOSFET at 120 kHz switching frequency while maintaining an 80% duty ratio operating point. Table 4 represents the parameter design specification of the modified circuit. The corresponding output voltage was taken for a duty ratio ranging from 0.1-0.9 with the maximum voltage drop being at k = 0.9. The reduction in voltage is due to an increase in the switching losses. It is a known fact that the extreme duty ratio affects the performance of the switch and also dwindles the practical efficiency of the converter [8, 9]. Figure 26 depicts a voltage gain curve of results obtained from the hardware prototype circuit and Lt-spice simulation result. It is observed that the voltage gain from both platforms is approximately 29V/V at k=0.9. A close observation indicates that both the Lt-spice and hardware prototype results conform with the theoretical investigation.

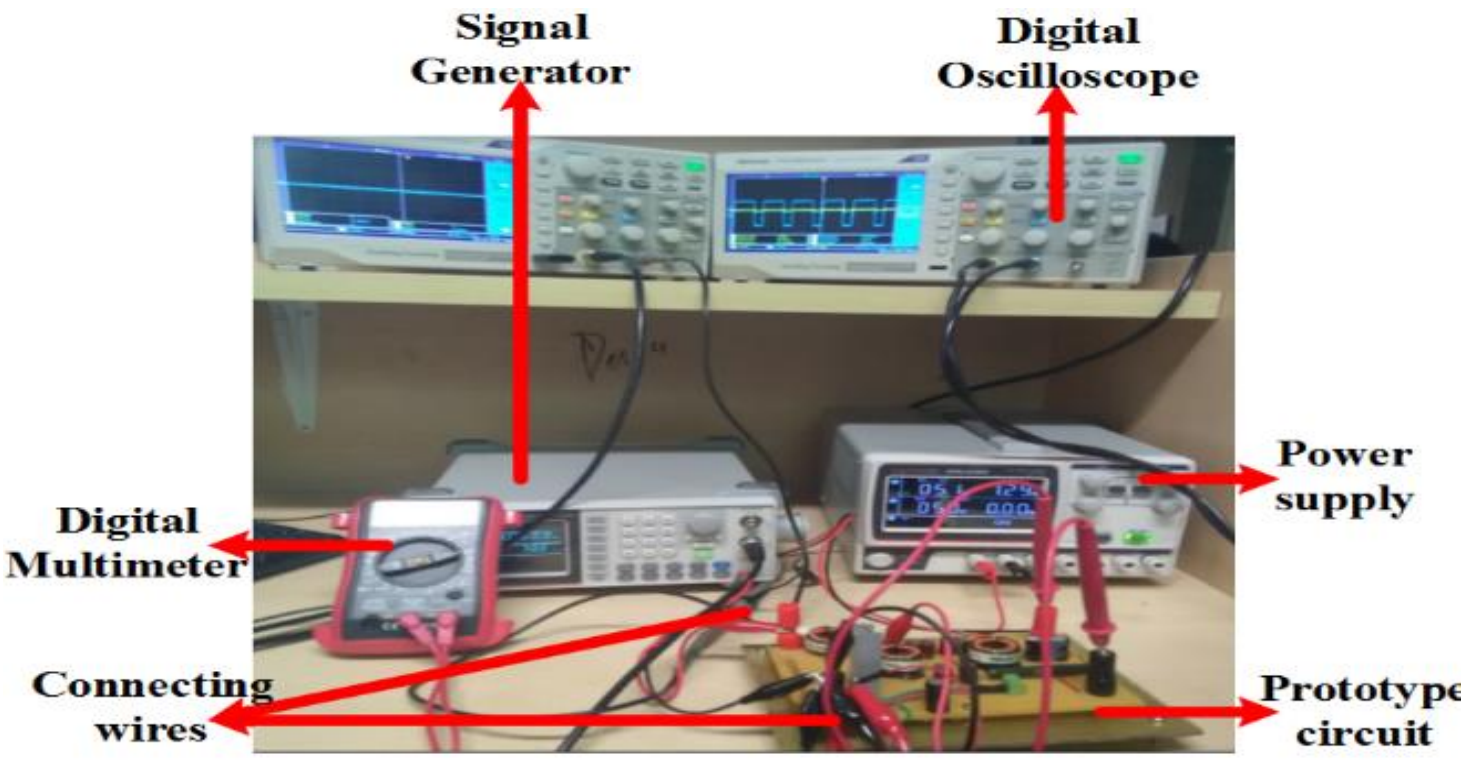

**Figure: 25**: Experimental Prototype Set-up of the Proposed Circuit.

**Table 4:** Experimental design specification of the proposed circuit

| PARAMETER | VALUE |
|---|---|
| Input voltage | 5V |
| Output power | 0.35Watt |
| Switching frequency | 100kHz |
| Output voltage | 70V |
| Resistor | 1000Ω |

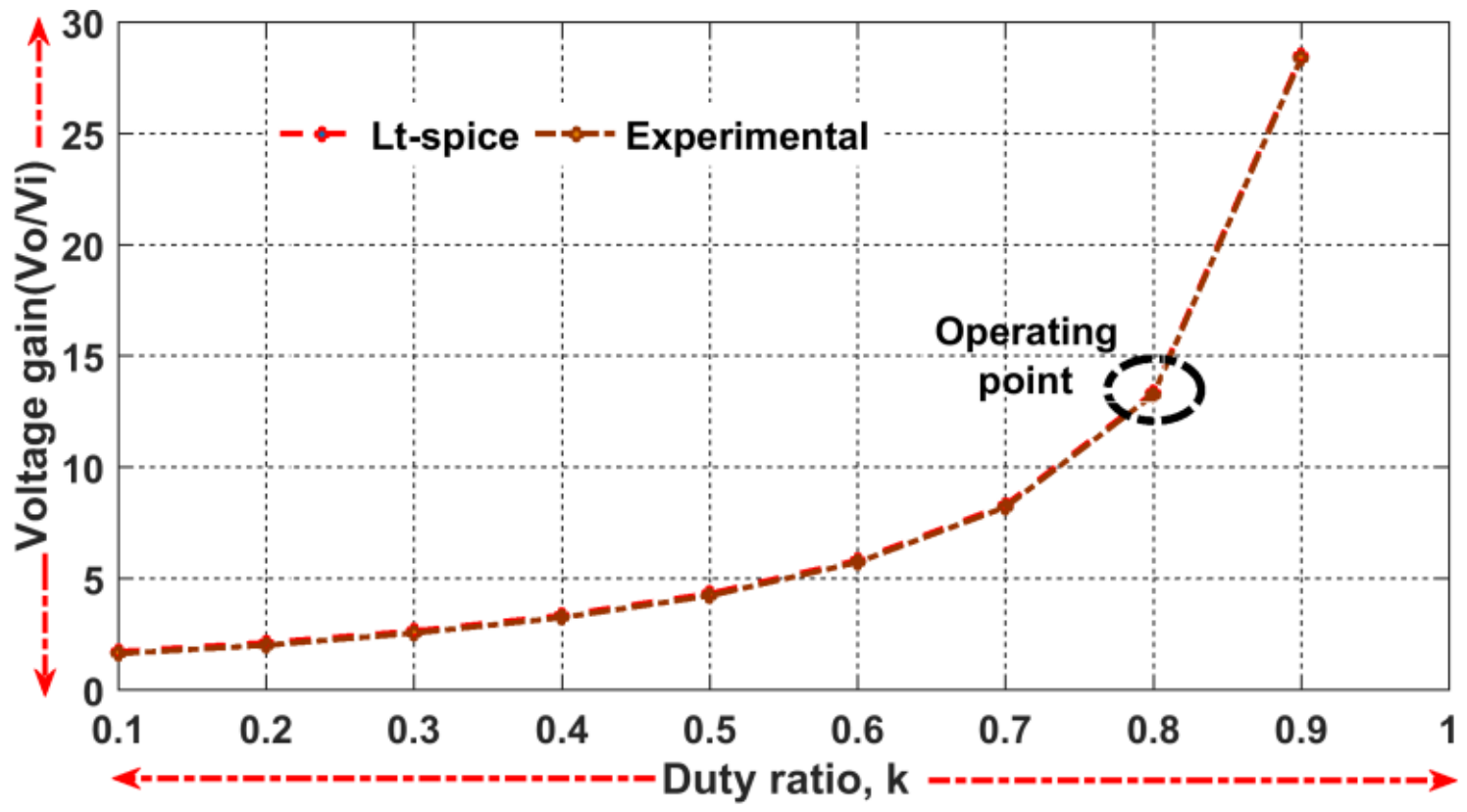

**Figure. 26**: Duty ratio comparison of proposed modified converter

As further verification of the converter performance accuracy, a close comparison is drawn between the proposed converter's experimental prototype waveforms and Lt-spice simulation performances for all the capacitors and diodes components. The waveforms are presented as shown in Fig. (27)-Fig. (35) below. Starting with the experimental results, the yellow line represents the gate signal, $V_{Gs}$ while the blue color illustrates the component waveforms. In the same manner, red and green color in Lt-spice simulation waveforms illustrates the gate signal, $V_{Gs}$ and component waveforms respectively.

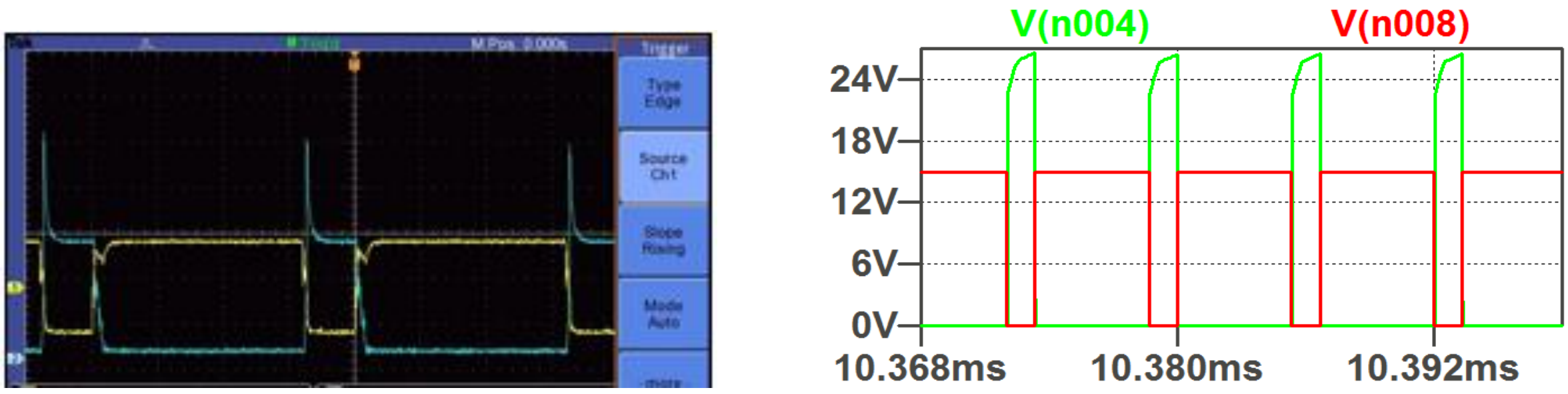

**(a)** **(b)**

**Figure 27:** Gate signal, $V_{GS}$ and $V_{DS}$ (a) prototype waveform (b) Lt-spice waveform.

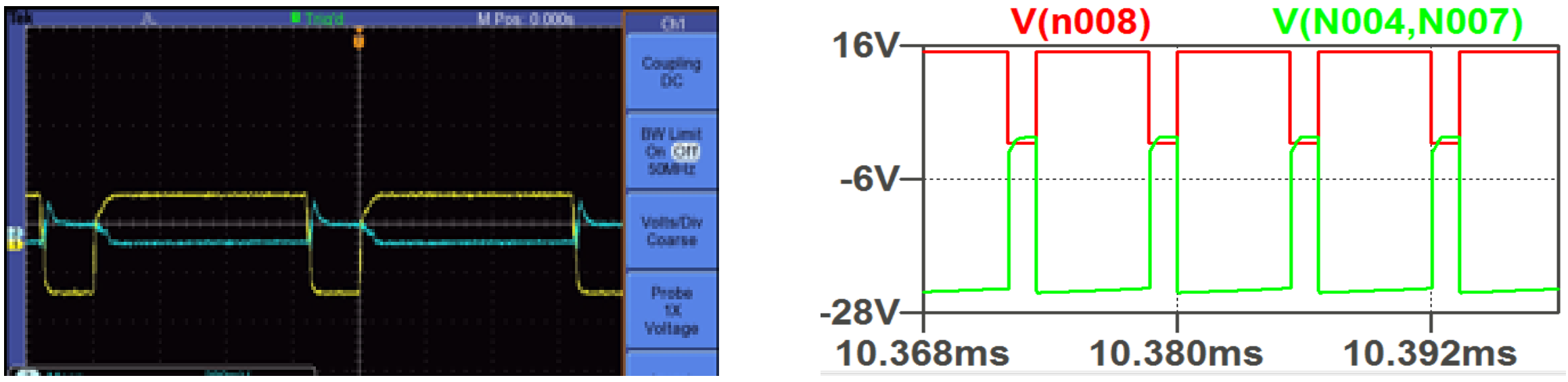

(a) (b)

**Figure 28:** Diode voltage, $V_{D1}$ (a) prototype waveform (b) Lt-spice waveform.

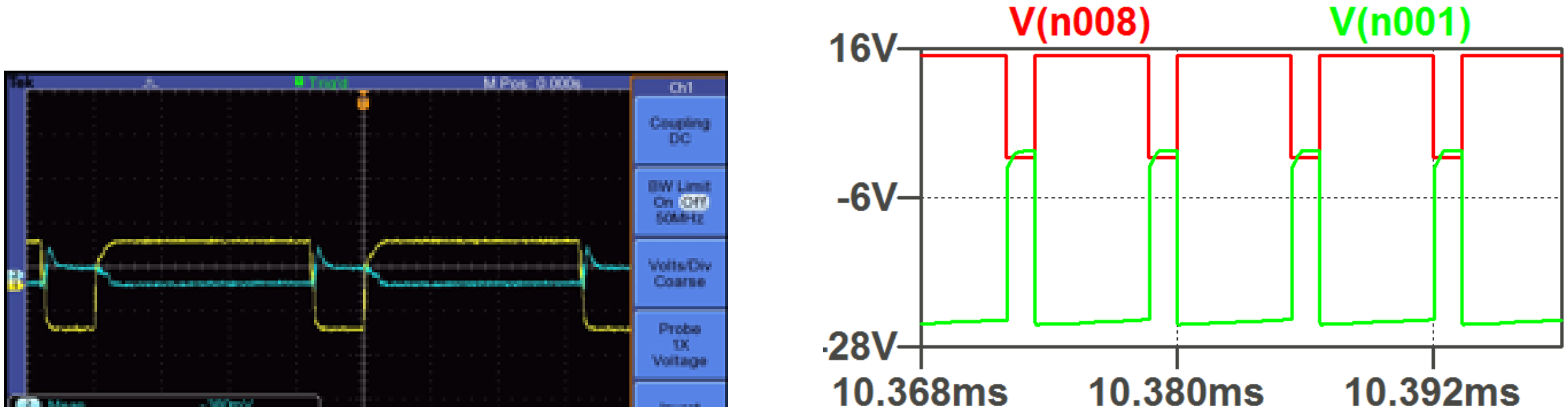

(a) (b)

**Figure 29:** Diode voltage, $V_{D2}$ (a) prototype waveform (b) Lt-spice waveform.

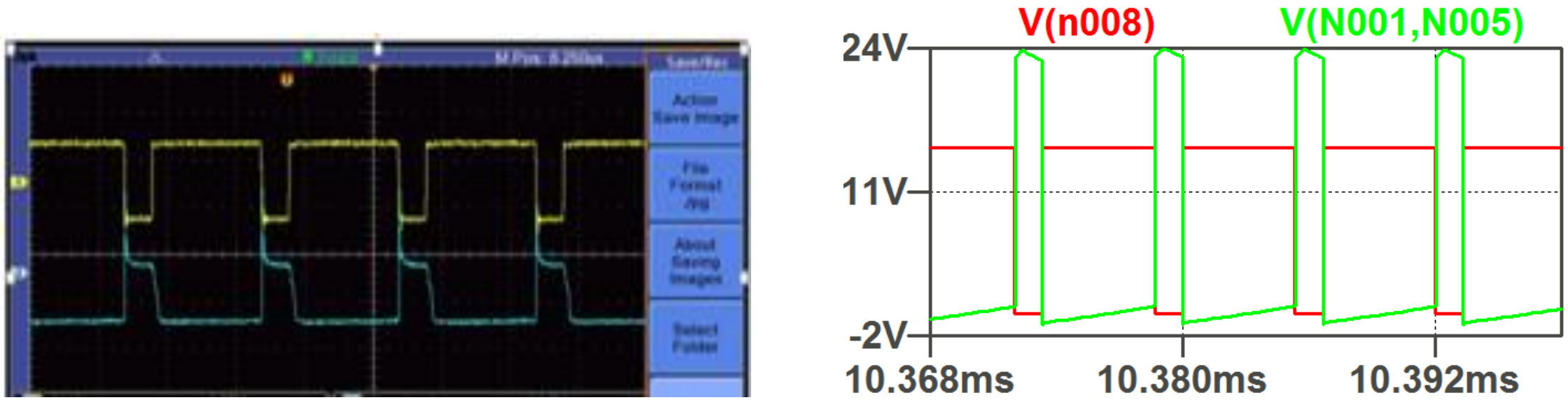

(a) (b)

**Figure 30:** Diode voltage, $V_{D3}$ (a) prototype waveform (b) Lt-spice waveform.

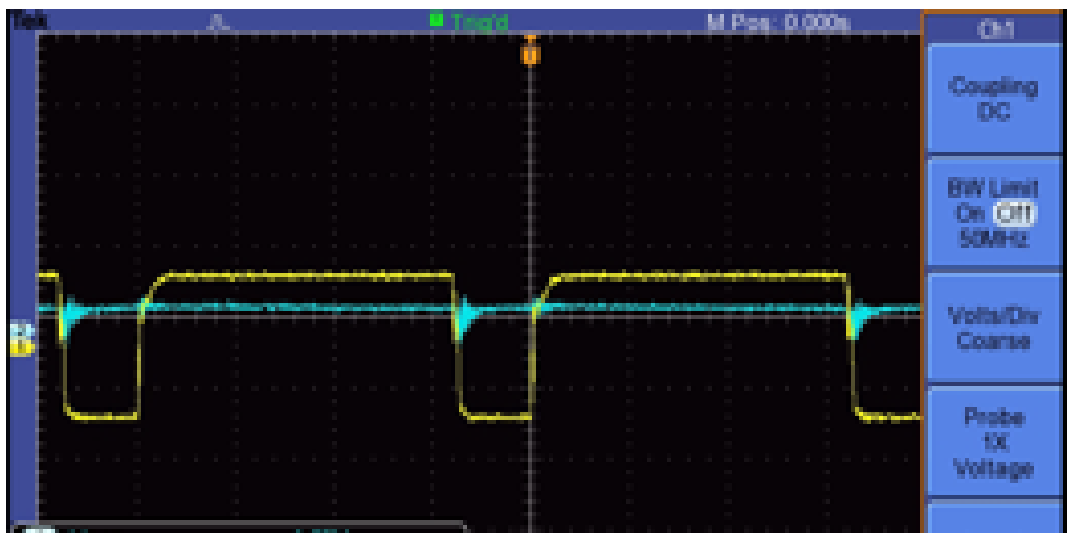

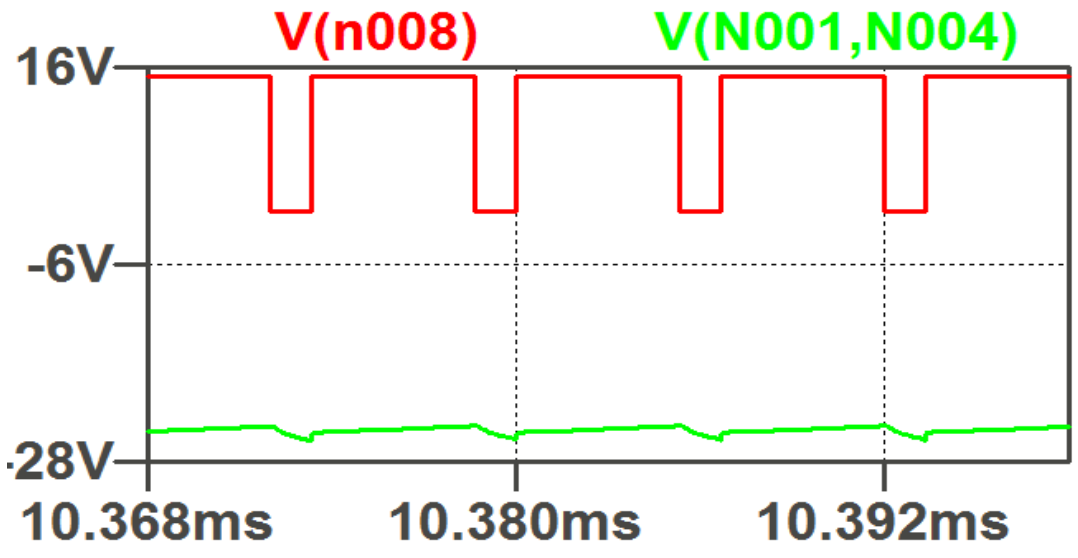

(a) (b)

**Figure 31:** Capacitor Voltage, $V_{C1}$ (a) prototype waveform (b) Lt-spice waveform.

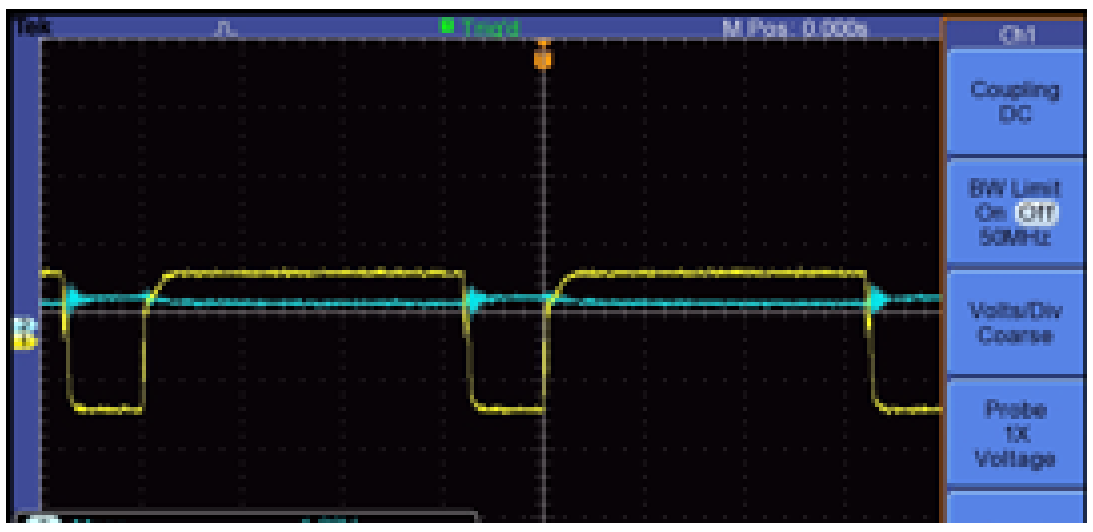

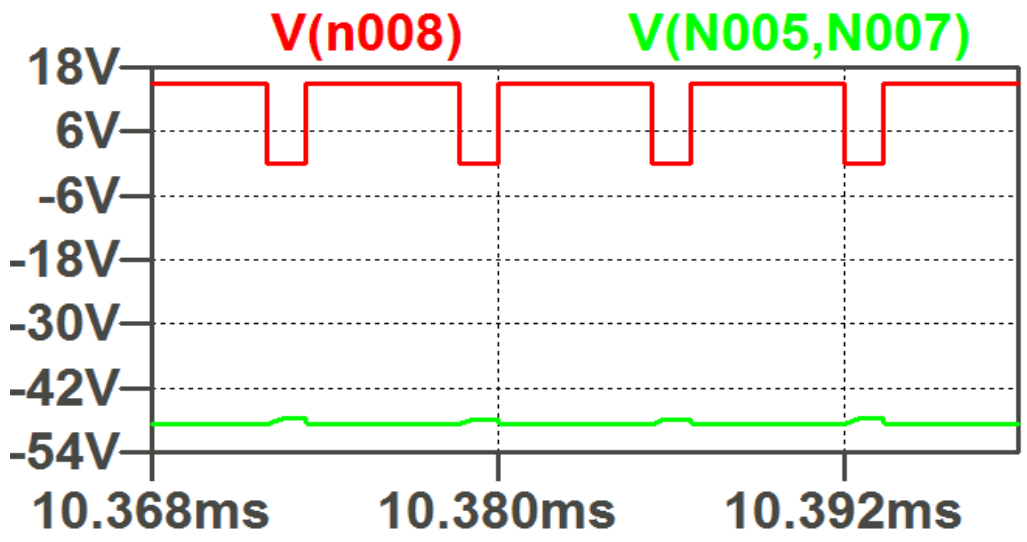

(a) (b)

**Figure 32:** Capacitor voltage, $V_{C2}$ (a) prototype waveform (b) Lt-spice waveform.

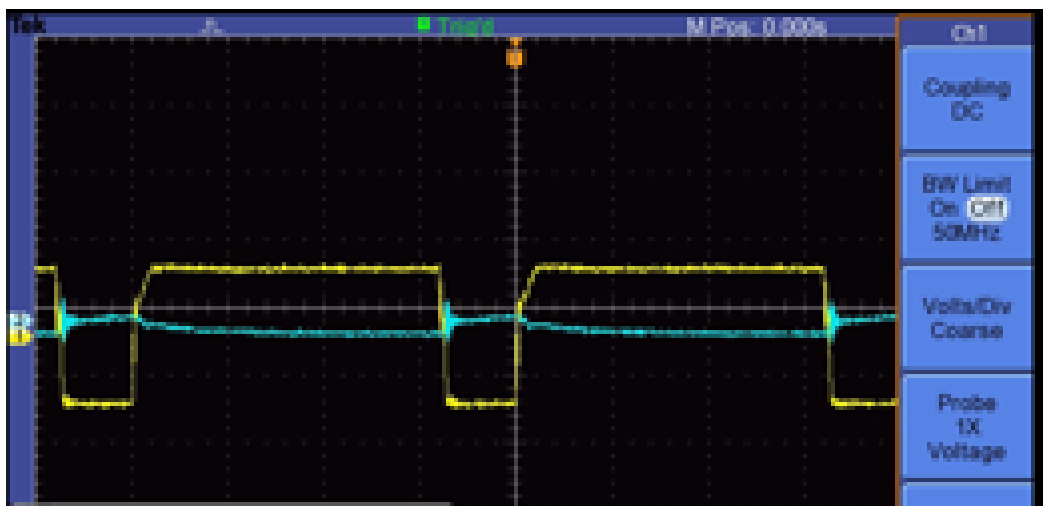

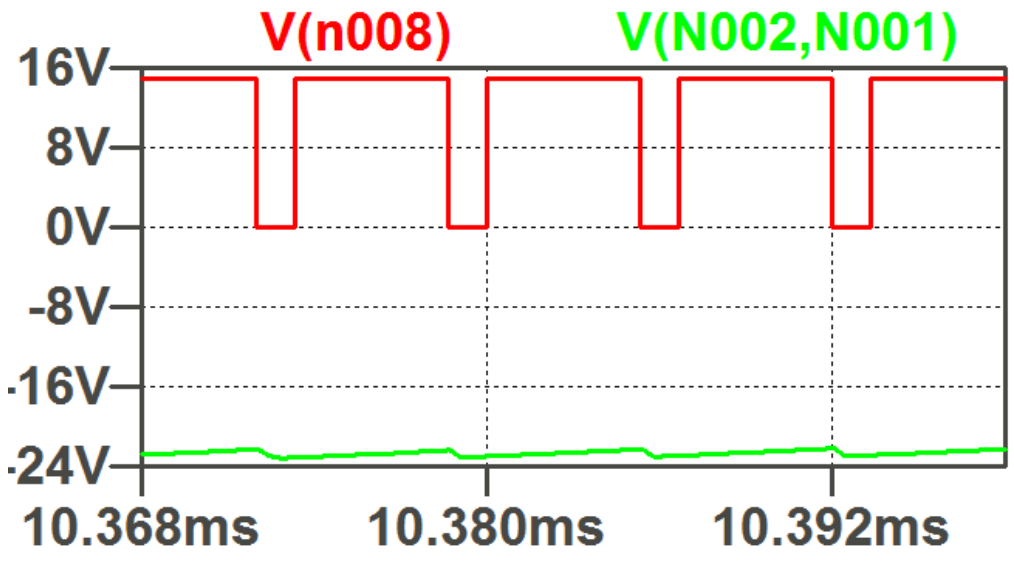

(a) (b)

**Figure 33:** Capacitor voltage, $V_{C3}$ (a) prototype waveform (b) Lt-spice waveform.

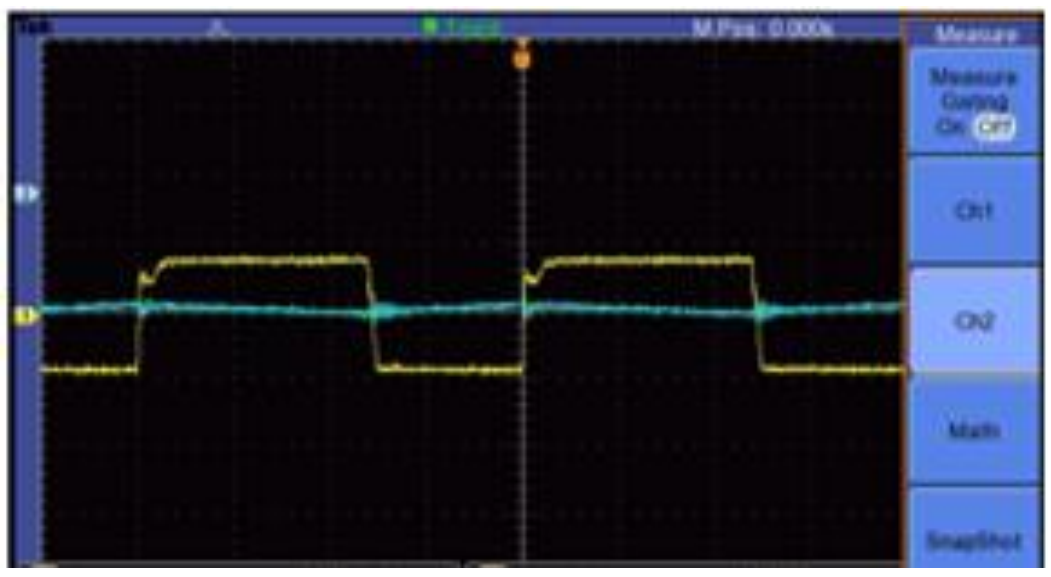

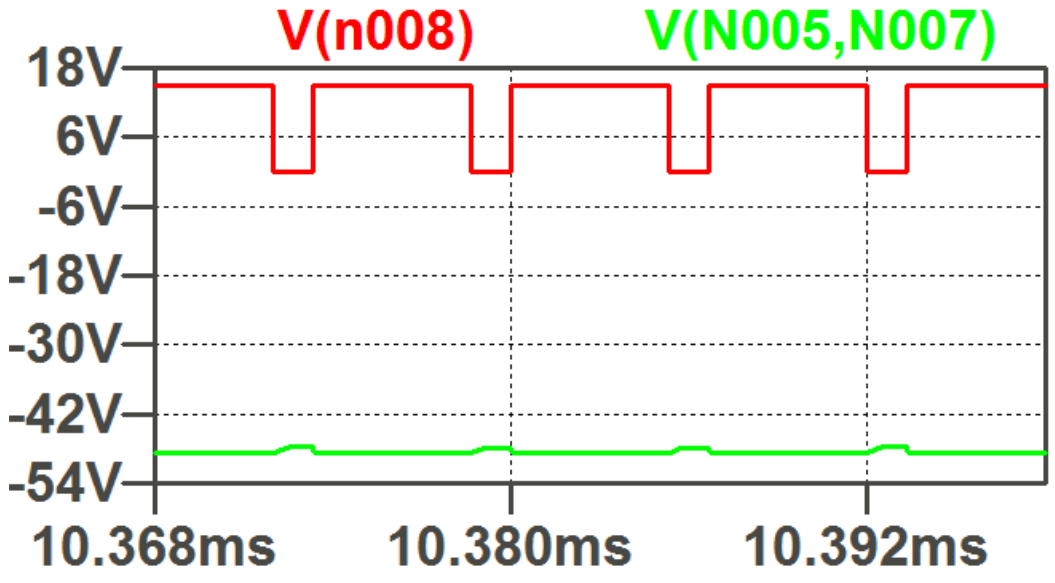

(a) (b)

**Figure 34:** Capacitor voltage, $V_{C4}$ (a) prototype waveform (b) Lt-spice waveform.

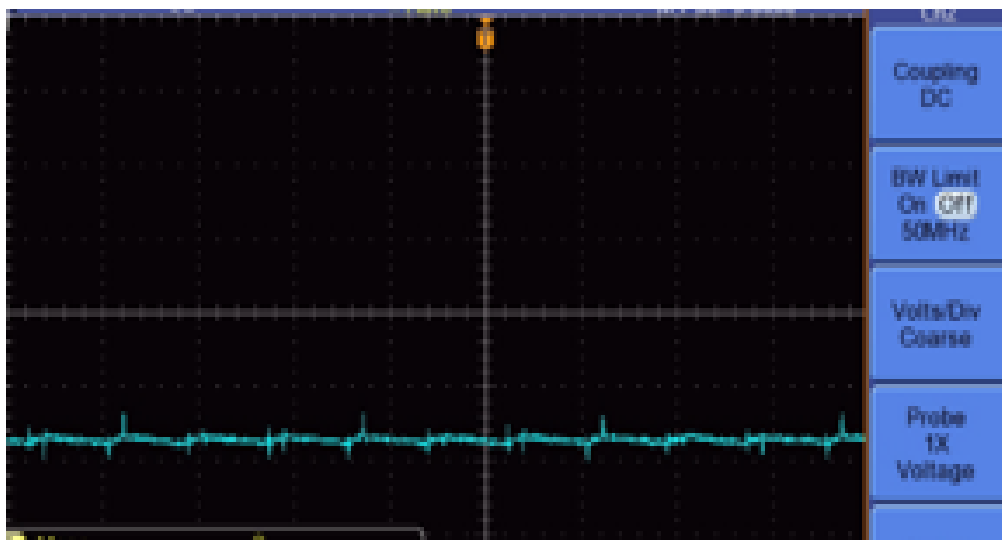

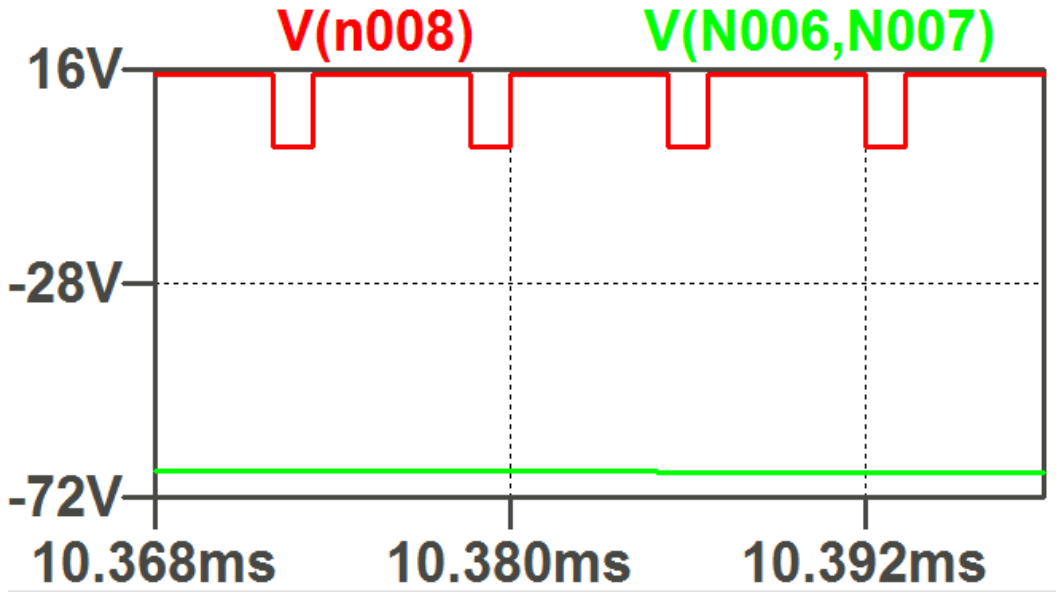

(a) (b)

**Figure 35:** Output Capacitor voltage, $V_{Co}$ (a) prototype waveform (b) Lt-spice waveform.

## 8. Conclusions

Three high-gain modified Cuk converters were proposed in this paper. The circuits generate higher voltage gain with fewer number of components compared to similar circuits in the literature. In perspective view, the first, second, and third proposed converters respectively accomplished voltage gains of 10V/V, 20V/V, and 29V/V at 90% duty ratio. The theoretical investigation confirmed the above values. Lt-spice and experimental prototype results of the first, second, and third proposed modified circuit were explicitly reported. In each case, the results show conformity to the theoretical analysis. A lower component count implies a reduction in production and this also enhances portability. The major drawback associated with the three proposed circuits is a reduction in efficiency at a 90% duty ratio. The switching losses are observed to increase at a higher value of k leading to a large voltage drop. This makes operating at peak duty ratio very challenging.

## References.

**Acknowledgments:** No source of funding for this project.

**Author Contributions**